\documentclass[twocolumn,preprintnumbers,superscriptaddress
,nofootinbib %footinbib for PRL
]{revtex4-1}

\usepackage{simplewick}

\newcommand{\beq}{\begin{eqnarray}}
\newcommand{\eeq}{\end{eqnarray}}
\newcommand{\e}{\mathrm{e}}

\newcommand{\im}{\mathrm{i}}
\newcommand{\p}{\partial}
\newcommand{\calO}{\mathcal{O}}

\usepackage{graphicx}
\usepackage{amssymb}
\usepackage{amsmath}
\usepackage{color}
\usepackage{epsfig}
\usepackage{epstopdf}
\usepackage{hyperref}
\usepackage{subfigure}
\allowdisplaybreaks

\usepackage{qcircuit}
\usepackage{mathtools}

\DeclarePairedDelimiterX\braket[2]{\langle}{\rangle}{#1 \delimsize\vert #2}

\begin{document}

\preprint{YITP-21-12, UT-Komaba-21-4}
\date{\today}
%\title{Digital quantum simulation for 
%screening and confinement \\
%in gauge theory with  a topological term}
\title{{Classically emulated digital quantum simulation for screening and confinement in the Schwinger model with a topological term}}
\author{Masazumi Honda}
\email[]{masazumi.honda(at)yukawa.kyoto-u.ac.jp}
\affiliation{{\it
Center for Gravitational Physics, Yukawa Institute for Theoretical Physics,
Kyoto University, Sakyo-ku, Kyoto 606-8502, Japan
}}
\author{Etsuko Itou}
\email[]{itou(at)yukawa.kyoto-u.ac.jp}
\affiliation{{\it
Strangeness Nuclear Physics Laboratory,
RIKEN Nishina Center, Wako 351-0198, Japan
}}
\affiliation{{\it
Keio University, 4-1-1 Hiyoshi, Yokohama, Kanagawa 223-8521, Japan
}}
\affiliation{{\it
Research Center for Nuclear Physics (RCNP), Osaka University, Osaka 567-0047, Japan
}}

\author{Yuta Kikuchi}
\email[]{ykikuchi(at)bnl.gov}
\affiliation{{\it
Department of Physics, Brookhaven National Laboratory, Upton, NY, 11973, USA
}}

\author{Lento Nagano}
\email[]{lento(at)icepp.s.u-tokyo.ac.jp}
\affiliation{{\it
International Center for Elementary Particle Physics (ICEPP), The University of Tokyo, 7-3-1 Hongo, Bunkyo-ku, Tokyo 113-0033, Japan
}}

\author{Takuya Okuda}
\email[]{takuya(at)hep1.c.u-tokyo.ac.jp}
\affiliation{{\it
Graduate School of Arts and Sciences, University of Tokyo,
Komaba, Meguro-ku, Tokyo 153-8902, Japan
}}

%%%%%%%%%%%%%%%%%%%%%%%%%%%%%%%%%%%%%%
\begin{abstract}

We perform digital quantum simulation, using a classical simulator, to study screening and confinement in a gauge theory with a topological term,
focusing on ($1+1$)-dimensional quantum electrodynamics (Schwinger model) 
with a theta term.
We compute the ground state energy in the presence of probe charges to estimate the potential between them,
via adiabatic state preparation. 
We compare our simulation results 
and analytical predictions for a finite volume, finding good agreements.
In particular our result in the massive case shows a linear behavior for non-integer charges and a non-linear behavior for integer charges,
consistently with the expected confinement (screening) behavior for non-integer (integer) charges.
\end{abstract}
\maketitle

\renewcommand{\thefootnote}{\arabic{footnote}}
\setcounter{footnote}{0}

\newpage

%\tableofcontents %To be removed for PRL submission

\setcounter{page}{1}

%%%%%%%%%%%%%%%%%%%%%%%%%%%%%%%%%%%%%%
%%%%%%%%%%%%%%%%%%%%%%%%%%%%%%%%%%%%%%
%%%%%%%%%%%%%%%%%%%%%%%%%%%%%%%%%%%%%%
\section{Introduction}
%%%%%%%%%%%%%%%%%%%%%%%%%%%%%%%%%%%%%%
%%%%%%%%%%%%%%%%%%%%%%%%%%%%%%%%%%%%%%
%%%%%%%%%%%%%%%%%%%%%%%%%%%%%%%%%%%%%%
A historical milestone of the classical lattice simulation of quantum chromodynamics (QCD) 
was the derivation of the confinement potential from the first principles~\cite{Wilson:1974sk, Creutz:1980zw, Otto:1984qr}.
The potential energy~$V(R)$ between a quark and an antiquark was obtained, via the relation $\langle W(R,T) \rangle \sim \e^{- V(R)T }$ for 
large $T$,
by calculating the expectation value of the rectangular Wilson loop $W(R,T)$ with spatial distance~$R$ and temporal separation~$T$.
The resulting potential $V(R)$ can be fitted well~\cite{Bali:2000gf} by a function of the form $\sigma R- \alpha/R$ with $\sigma$ and $\alpha$ constants~\cite{Eichten:1974af}.
The linear growth of $V(R)$ for large $R$ is evidence, or the definition, of quark confinement, whose strength is quantified by
the string tension~$\sigma$.

The conventional lattice QCD is based on the Lagrangian formalism and the path integral is usually computed by the Markov chain Monte Carlo algorithm,
where the Euclidean action~$S$ defines the probability~$\e^{-S}$ assigned to a field configuration on discretized spacetime.
While very successful for real~$S$, the conventional lattice simulation becomes practically intractable 
when $S$ is complex and the integrand is highly oscillatory. 
This is called the sign problem and arises, for example, 
when we turn on the topological (theta) term in the action, introduce quark chemical potentials, or consider the real-time dynamics of QCD.
These situations occur in many important
subjects such as
the strong CP problem, neutron stars, the early universe and collider experiments.  
Various alternative methods have been proposed within the path integral formalism.  
See, for example,~\cite{Aarts:2015tyj}.

In {\it digital quantum simulation}, where the Hamiltonian formalism is favored 
and quantum operations over the exponentially large Hilbert space are expected to be efficiently processed, the sign problem tied to the path integral is absent from the beginning.
We still need to formulate the theory on a finite lattice and regularize the infinite-dimensional Hilbert space.
While the small number of qubits and high error rates severely limit the capability of current quantum computers,
it is hoped that quantum hardware with sufficient (supreme) abilities will be built in the near (far) future to allow digital simulation of quantum field theories.
Before such machines are realized, it is important to develop simulation methods and demonstrate their usefulness.
Even now we can use
a simulator
on a classical computer to perform quantum simulations with a moderate number of qubits {(up to $20$ or $30$ qubits depending on tasks)} and without quantum noise.

In this {work}
%letter, 
we perform digital quantum simulation on a classical simulator to study screening and confinement in a gauge theory with a topological term.
We focus on ($1+1$)-dimensional quantum electrodynamics (Schwinger model~\cite{Schwinger:1962tp}), 
whose Lagrangian with a theta term~\cite{Coleman:1975pw} is given by~\footnote{Our convention: $\eta_{\mu\nu}={\rm diag}(1,-1), \epsilon_{01}=1$, 
$F_{\mu\nu}=\partial_\mu A_\nu -\partial_\nu A_\mu$,
$\gamma^0 = \sigma_z, \gamma^1 =\im\sigma_y, \gamma^5 =\sigma_x$, $\bar\psi=\psi^\dagger\gamma^0$.
}
\begin{align}
\mathcal{L}_{\rm con}
&= -\frac{1}{4} F_{\mu\nu} F^{\mu\nu} +\frac{g\theta}{4\pi} \epsilon_{\mu\nu} F^{\mu\nu}
\nonumber\\
&\qquad
 +\im\bar{\psi}\gamma^\mu (\p_\mu +\im gA_\mu ) \psi -m\bar{\psi}\psi .
\label{eq:Lagrangian}
\end{align}
Here $\psi$ is the Dirac fermion, $g$ the gauge coupling, and $m$ the fermion mass.
The model provides a good testing ground for quantum algorithms  
as it is one of the simplest nontrivial gauge theories that share some features with QCD.
With two probe particles of charges~$\pm q$, the model is known to exhibit qualitatively different behaviors depending on the values of $q$ and $m$ (for $\theta=0$):
with $m=0$ the two probe charges are screened for any value of $q$~\cite{Gross:1995bp}, 
while with $m\neq  0$
they are confined for non-integer $q$ and screened for integer $q$~\cite{Coleman:1975pw}.

To study the screening versus confinement problem,
we compute the ground state energy in the presence of probe charges and
estimate the potential between them.
We put the Schwinger model on a lattice and introduce two probe charges $\pm q$ 
using a position-dependent theta angle.
Then we adiabatically prepare the 
ground state of the Hamiltonian and compute its energy as a function of a distance between the probes~\footnote{
{One can use an adiabatic process to prepare the ground state accurately for finite-gap systems by taking a sufficiently large adiabatic time. 
%We chose adiabatic state preparation because we expect that 
In the fault-tolerant era it is anticipated to be more useful than variational methods.
%it will be more useful than variational methods in the fault-tolerant era.
}
}.
This is an analog of the quark-antiquark potential
while our computational scheme should be contrasted with the one in the conventional lattice QCD.
%\red{Even with a modest number of qubits,}
Our simulation results in the massless case agree well with the analytic results for a finite volume
in the continuum limit.
In the massive case, 
our result shows a linear behavior for non-integer $q$
just like the confining potential in the infinite-volume limit does.
The present work envisions what the investigation will look like when one studies screening and confinement on a real quantum computer in the future.
We believe that our work will serve as a guide for the simulation of many-body physics in the noisy intermediate-scale quantum~(NISQ)~\cite{Preskill2018quantumcomputingin}/early fault-tolerant era, when the number of 
%physical (logical) 
qubits will be limited~\footnote{
Previous works that study digital quantum simulation of the Schwinger model include~{\cite{Martinez:2016yna,Muschik:2016tws,Klco:2018kyo,Kokail:2018eiw,Magnifico:2019kyj,Chakraborty:2020uhf,Yamamoto:2021vxp,2020PhRvR...2b3342K,2021arXiv210608394D}}.
See also~\cite{Bernien_2017,Surace_2020} for analog simulations of the Schwinger model, \cite{Jordan:2011ne,Jordan:2011ci,Jordan:2014tma,Garcia-Alvarez:2014uda,Wiese:2014rla,Marcos:2014lda,Mezzacapo:2015bra,Macridin:2018gdw,Lamm:2018siq,Klco:2018zqz,Gustafson:2019mpk,Alexandru:2019ozf,Klco:2019xro,Klco:2019evd,Lamm:2019uyc,Mueller:2019qqj,Gustafson:2019vsd,2020arXiv200615746B,2020PhRvL.125p0503M,2020arXiv201007965A,2020arXiv201106576B,2020PhRvA.102e2422K,2020JHEP...12..011L,2020arXiv201209194S,2021PhRvD.103e4507G,2021PhRvA.103d2410B,2021PhRvD.103i4501C,2021JHEP...07..140G,2021PhRvD.104a4512E,2021PRXQ....2c0334P,2021PhRvD.104h6013L,2021PhRvD.104g4505D,2021PhRvL.127u2001B}
for digital simulations of other quantum field theories, and \cite{Zohar:2012ay,Banerjee:2012pg,Zohar:2012xf,Banerjee:2012xg,Wiese:2013uua,Zohar:2015hwa,Bazavov:2015kka,Zohar:2016iic,Bermudez:2017yrq,Zache:2018jbt,Zhang:2018ufj,Lu:2018pjk,Roy:2020ppa} for their analog simulations.
{Algorithms specific to the simulation of gauge theories were developed in~\cite{2020arXiv200616248T,2021PhRvD.103i4501C,2020PhRvD.101k4502R,2021PhRvD.103k4505G}.}
}.

%%%%%%%%%%%%%%%%%%%%%%%%%%%%%%%
%%%%%%%%%%%%%%%%%%%%%%%%%%%%%%%
%%%%%%%%%%%%%%%%%%%%%%%%%%%%%%%
\section{Qubit description of the Schwinger model}
\label{sec:qubit-Schwinger}

We begin by rewriting the Hamiltonian of the Schwinger model in terms of spin operators acting on qubits, 
following~\cite{Banks:1975gq,Hamer:1997dx,Martinez:2016yna} but with a theta angle~\cite{Chakraborty:2020uhf}.
The Hamiltonian of the continuum theory in the gauge $A_0=0$ is~\footnote{In contrast to~\cite{Chakraborty:2020uhf}, we do not absorb the theta angle by a chiral rotation of the mass term.} 
\begin{align}
\int dx \left[ \frac{1}{2} \Big( \Pi -\frac{g\theta}{2\pi} \right)^2
-\im \bar{\psi} \gamma^1 (\p_1 +\im g A_1) \psi +m\bar{\psi} \psi 
\Big],
\nonumber
\end{align}
where $\Pi=\partial_0 A^1 +g\theta/2\pi$  
is the canonical momentum conjugate to $A^1$ and
physical states are subject to the Gauss law $\p_1 \Pi +g\psi^\dag \psi =0$.

We regularize the theory by putting it on a one-dimensional spatial lattice 
with $N$ sites, lattice spacing~$a$ and open boundaries.
We replace the Dirac fermion
$\psi(x)=\big(\psi_u(x),\psi_d(x) \big){}^\mathsf{T}$
by a pair of neighboring
staggered fermions~\cite{Kogut:1974ag}
according to
\begin{align}
\frac{\chi_n}{\sqrt{a}} \ \longleftrightarrow \ \left\{ 
\begin{array}{ll} 
\psi_u (x) & n:{\rm even}\,, \cr
\psi_d (x) & n:{\rm odd}\,, \end{array}
\right. 
\end{align}
where $n=0,\ldots,N-1$ label the lattice sites $x=na$.
The lattice counterparts of the gauge field operators are 
$U_n$ ($\leftrightarrow$ $\e^{-\im ag A^1 (x)}$) and $L_n$ ($\leftrightarrow$ $-\Pi(x)/g$) such that $U_n^\dagger=U_n^{-1}$, $L_n^\dagger=L_n$,
defined on the link between the $n$-th and $(n+1)$-th sites.
These operators satisfy the canonical commutation relations
\begin{align}
\{ \chi_n^\dag ,\chi_m \} = \delta_{mn},\  \{ \chi_n ,\chi_m \} =0 ,\   
[U_n ,L_m] =\delta_{mn}U_n .
\nonumber
\end{align}

Let us introduce the probe charges $+q$ and  $-q$ on the $\hat{\ell}_0$-th and $(\hat{\ell}_0+\hat{\ell})$-th sites, respectively.
We realize this by making the theta angle position dependent as
\begin{align}
\label{eq:theta_config}
\vartheta_n 
= \left\{
 \begin{array}{cc}
  2\pi q +\theta_0, & \hat{\ell}_0 \le n < \hat{\ell}_0+\hat{\ell},\\
  \theta_0, & \text{otherwise}.
 \end{array}
 \right.
\end{align}
We place the probe charges equally separated from the 
center of the lattice 
by setting $\hat{\ell}_0=(N-1-\hat{\ell})/2$.
The lattice Hamiltonian 
with the probe charges is
\begin{align}
&H 
= J \sum_{n=0}^{N-2} \left( L_n +\frac{\vartheta_n}{2\pi}  \right)^2 
 -\im w \sum_{n=0}^{N-2} \big( \chi_n^\dag U_n \chi_{n+1} 
  \nonumber
 \\
 &\qquad\qquad
 -\chi_{n+1}^\dag U_n^\dag \chi_{n} \big)  +m\sum_{n=0}^{N-1} (-1)^n \chi_n^\dag \chi_n,
 \label{eq:fermion-Hamiltonian}
\end{align}
where $w=1/(2a)$ and $J=g^2a/2$.
The physical states must satisfy the lattice version of the Gauss law
\begin{align}
L_n -L_{n-1} =  \chi_n^\dag \chi_n -\frac{1-(-1)^n}{2}.
\label{eq:Gauss_lattice}
\end{align}
We solve this
with the boundary condition
$L_{-1}=0$ and fix the gauge $U_n =1$
to eliminate $(L_n ,U_n )$.

To rewrite the theory in terms of a spin system, we apply
the Jordan-Wigner transformation~\cite{Jordan1928}
\begin{equation}
 \chi_n = \frac{X_n-\im Y_n}{2}\prod_{i=0}^{n-1}(-\im Z_i), 
\end{equation}
where ($X_n, Y_n ,Z_n$) denote the Pauli matrices ($\sigma_x$, $\sigma_y, \sigma_z$)
acting on the qubit on site~$n$,
then the lattice Hamiltonian becomes~\cite{Kogut:1974ag}
\begin{align}
 H &= J\sum_{n=0}^{N-2} \left[\sum_{i=0}^{n}\frac{Z_i + (-1)^i}{2}+\frac{\vartheta_n}{2\pi}\right]^2 
 \nonumber
 \\
 &
 + \frac{w}{2}\sum_{n=0}^{N-2}\big[X_n X_{n+1}+Y_{n}Y_{n+1}\big]
 + \frac{m}{2}\sum_{n=0}^{N-1}(-1)^n Z_n ,
\label{eq:Hspin_theta}
\end{align}
up to an irrelevant constant.
In this form, it is manifest that we can directly apply quantum algorithms {in qubit form} to the lattice Schwinger model.

%%%%%%%%%%%%%%%%%%%%%%%%%%%%%%%
%%%%%%%%%%%%%%%%%%%%%%%%%%%%%%%
%%%%%%%%%%%%%%%%%%%%%%%%%%%%%%%
\section{Simulation protocol}
\label{sec:protocol}
%%%%%%%%%%%%%%%%%%%%%%%%%%%%%%%
%%%%%%%%%%%%%%%%%%%%%%%%%%%%%%%
%%%%%%%%%%%%%%%%%%%%%%%%%%%%%%%
Our main target is the energy
\beq
\label{eq:energy-def}
E(\theta_0,q,\ell) := 
\langle {\rm GS}| H | {\rm GS} \rangle = \langle H  \rangle 
\eeq
of the ground state $|{\rm GS}\rangle$ in the presence of probes, where the parameters~$(\theta_0,q,\ell=\hat{\ell}a)$ enter the setup 
as described above.
We will refer to~$V_f(\theta_0,q,\ell):= E(\theta_0,q,\ell)-E(0,0,0)$ as the potential~\footnote{
The $\ell$-dependent part 
%of the energy 
is free from UV divergence and
therefore the potential has a well-defined continuum limit.}.
We use the adiabatic state preparation and the Suzuki-Trotter decomposition to obtain an approximation to the ground state $|{\rm GS} \rangle$.

Set $H_0:=H|_{w=0,\theta_0 =0, q=0,m=m_0}$ for some $m_0>0$.
Its ground state is 
$|{\rm GS}_0 \rangle =| 1010\dots \rangle$ with $Z|0\rangle = +|0 \rangle$ and $Z|1\rangle = -|1\rangle$, 
and can be constructed easily. 
Fix $T>0$ and choose a one-parameter family of slowly varying Hamiltonians $H_A(t)\ (0\leq t\leq T)$ such that
\begin{equation}
    H_A(0)=H_0, \qquad  H_A (T)=H.
\end{equation}
By the adiabatic theorem, if $H_A (t)$ has a unique gapped ground state for any~$t$, then the exact ground state $|{\rm GS}\rangle$ is given by 
\begin{align}
|{\rm GS}\rangle 
= \lim_{T\rightarrow\infty} \mathcal{T}\exp{\left(-{\rm i} \int_0^T dt\ H_A (t) \right)} 
|{\rm GS}_0 \rangle ,
\label{eq:evolution}
\end{align}
where $\mathcal{T}$ denotes time ordering.
In practice, we take $T$ to be sufficiently large but finite, which induces a systematic error in the preparation of the ground state (see, {\it e.g.},~\cite{farhi2000quantum,Jansen2007} and Appendix~\ref{sec:sys-error}).
%and Appendix~\ref{sec:sys-error}).

To apply the Suzuki-Trotter decomposition, we decompose the Hamiltonian as
\beq \label{eq:H_decomposition0}
 H = H_{XY}^{(0)} + H_{XY}^{(1)} + H_{Z} + C .
\eeq
Each term
is given as
\begin{widetext}
\begin{align}
\begin{split}
 &H_{XY}^{(0)} = \frac{w}{2} \sum_{m=0}^{\frac{N-3}{2}}\big(X_{2m}X_{2m+1}+Y_{2m}Y_{2m+1}\big),
 \qquad
 H_{XY}^{(1)} = \frac{w}{2} \sum_{m=1}^{\frac{N-1}{2}}\big(X_{2m-1}X_{2m}+Y_{2m-1}Y_{2m}\big),
 \\
  &H_{Z} = \frac{J}{2}\sum_{n=0}^{N-3} \sum_{m=n+1}^{N-2}(N-m-1)Z_n Z_m
  +\frac{J}{2}\sum_{n=0}^{N-2}
  \frac{1+(-1)^n}{2}
 \sum_{i=0}^{n}Z_i
 +qJ
\sum_{m=0}^{\hat\ell_0+\hat\ell-1}
(\hat\ell_0+\hat\ell-m)Z_m
 \\
&
\qquad\qquad
-qJ 
\sum_{m=0}^{\hat\ell_0-1}(\hat\ell_0-m)Z_m
+\frac{\theta_0}{2\pi}J\sum_{m=0}^{N-2}(N-m-1)Z_m+ \frac{m}{2}\sum_{n=0}^{N-1}(-1)^n Z_n,
\\
&C =
\frac{qJ}{2}\left(\hat\ell+(-1)^{\hat\ell_0}\frac{1-(-1)^{\hat\ell}}{2}\right)
+
q\left(q+\frac{\theta_0}{\pi}\right)J\hat\ell
+\frac{\theta_0J}{4\pi}\left(N-1+\frac{1 +(-1)^N}{2}\right)
+ \left(\frac{\theta_0}{2\pi}\right)^2J(N-1)
\end{split}
\label{eq:H_decomposition} 
\end{align}
\end{widetext}
for odd $N$, on which we focus in this work~\footnote{{In these expressions, only $X_n,Y_n,Z_n$ are quantum operators and all the other quantities are c-numbers.
%operators are not.
}}.
{Here $C$ is a classical number 
where we have kept only the terms that depend on $\theta_0,q,\hat{\ell}$.}
All the terms commute within each of $H_{XY}^{(0)}, H_{XY}^{(1)}$, and $H_Z$.

Let $H_{XY,s}^{(0)}, H_{XY,s}^{(1)}$, and $H_{Z,s}$ denote the modifications of the corresponding operators in~(\ref{eq:H_decomposition}) by the replacements
\begin{align}
 &w\to w\frac{s \delta t}{T}, \quad
 \theta_0\to \theta_0\frac{s \delta t}{T}, \quad
 q\to q\frac{s \delta t}{T}, 
 \nonumber
 \\
 &
 m\to m_0 \left(1-\frac{s \delta t}{T}\right) + m\frac{s \delta t}{T}.
 \label{eq:param-adiabatic}
 \end{align}
We approximate~(\ref{eq:evolution}) by
the second-order Suzuki-Trotter decomposition as
\begin{align}
\label{eq:SK_timeDep}
 |{\rm GS}_A \rangle \ :=
 & \
\prod_{s=1}^{M} \Big(
 \e^{-\im H_{XY,s}^{(0)}\frac{\delta t}{2}}\e^{-\im H_{XY,s}^{(1)}\frac{\delta t}{2}}
 \e^{-\im H_{Z,s}\delta t}
 \nonumber
 \\
 &\quad
 \times
 \e^{-\im H_{XY,s}^{(1)}\frac{\delta t}{2}}\e^{-\im H_{XY,s}^{(0)}\frac{\delta t}{2}}
 \Big)
|{\rm GS}_0 \rangle\ ,
\end{align}
where $M:=T/\delta t$ and the product is ordered from right to left with increasing $s$.
For fixed $T$, this approximation results in the error of $\calO(\delta t^2)$ for the whole operator~\cite{Lloyd1073,Suzuki1991}.
Then we compute $\langle {\rm GS}_A |H|{\rm GS}_A \rangle$, which approximates the ground state energy.
This involves separate measurements of the mutually non-commuting terms in the Hamiltonian~(\ref{eq:H_decomposition0}): $H_{XY}^{(0)}, H_{XY}^{(1)}$ and $ H_{Z}$.
We denote the number of 
times the circuit for each term is executed
by $n_\text{shots}$,
which 
determines the
statistical uncertainties in the simulation.
Details on the above protocol are explained in Appendix~\ref{app:Q_circuit}.

Let us comment on a property of the state $|{\rm GS}_A \rangle$.
The time evolution in~(\ref{eq:SK_timeDep}) preserves the ${\rm U}(1)$ symmetry generated by $Q:=\sum_{n=0}^{N-1} Z_n$
and the
state $|{\rm GS}_A \rangle$ approximates the ground state within the $Q=-1$ sector.
In principle, there
may be states with lower energies with different charges.
See Appendices~\ref{sec:sys-error} and~\ref{app:parity}  for details.

%%%%%%%%%%%%%%%%%%%%%%%%%%%%%%%%
%%%%%%%%%%%%%%%%%%%%%%%%%%%%%%%
%%%%%%%%%%%%%%%%%%%%%%%%%%%%%%%
%%%%%%%%%%%%%%%%%%%%%%%%%%%%%%%
\section{Simulation results}\label{sec:results}
%%%%%%%%%%%%%%%%%%%%%%%%%%%%%%%
%%%%%%%%%%%%%%%%%%%%%%%%%%%%%%%
%%%%%%%%%%%%%%%%%%%%%%%%%%%%%%%
We present the results of our quantum simulations on a classical simulator (QasmSimulator in IBM's open source SDK Qiskit).
In this section we fix $a=0.4 g^{-1}, \delta t =0.3g^{-1}, T=99g^{-1}$ and $m_0 =0.5g$.
Therefore the physical volume $L= a(N-1)$ is simply specified by $N$. 
In Appendix~\ref{sec:sys-error}
we show that the systematic errors are comparable to the statistical uncertainties in the whole parameter region we study while the systematic errors become larger  
when we expect a 
larger
effect of a phase transition~\footnote{
The continuum Schwinger model in the infinite volume undergoes a first-order phase transition at $\theta_0=\pi$ and for $m$ larger than a critical mass $ m_c$~\cite{Coleman:1976uz}, with value $ m_c/g  = 0.3335(2)$~\cite{Byrnes:2002nv}.
},
{\it i.e.,} for larger $\theta_0, m, q$, and $\ell$.
Analytical results in a finite volume compared here are derived in
Appendix~\ref{app:bosonization}.  

%%%%%%%%%%%%%%%%
\begin{figure*}[t]
\centering
\includegraphics[scale=0.75]{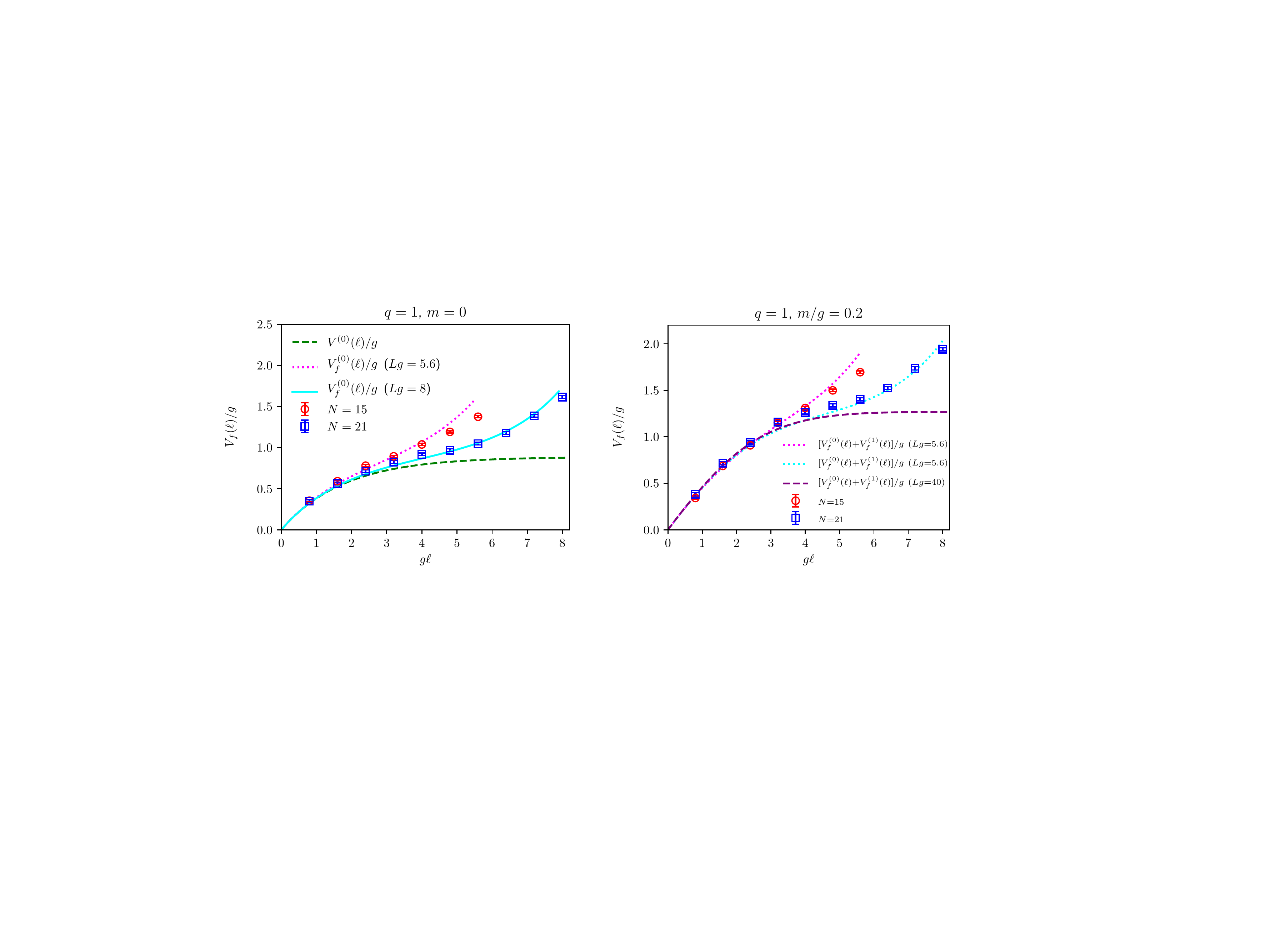}
\caption{
Results for
$V_f(\ell)$
with
$n_\text{shots}=10^5$.
The statistical uncertainties 
are 
smaller than the markers.
The curves (in colors similar to and lighter than the data markers) represent $V_f^{(0)}(\ell)$ on the left and $V_f^{(0)}(\ell)+V_f^{(1)}(\ell)$ on the right  with the corresponding volumes. The dashed curve (purple) on the right panel depicts $V_f^{(0)}(\ell)+V_f^{(1)}(\ell)$ for $N=101${, $Lg=40$}.  
}
\label{fig:potential-intQ}
\end{figure*}
%%%%%%%%%%%%%%%%

We begin with the $\theta_0=0$ case.
In FIG.~\ref{fig:potential-intQ}, 
we show the simulation results for the potential~$V_f$ with $q=1$ as a representative of the integer $q$ case
where the theory is known to exhibit a screening behavior.
For $m=0$ (left panel),
we compare the simulation results with the potential
in the infinite volume~\cite{Iso:1988zi}
\beq
V^{(0)}(\ell) 
= \frac{\sqrt{\pi}q^2 g}{2} \left( 1-\e^{-\frac{g \ell}{\sqrt{\pi}} } \right) ,
\label{eq:Gross-massless}
\eeq 
and the one in a finite volume
\beq
V_f^{(0)}(\ell) 
= \frac{\sqrt{\pi}q^2 g}{2} 
\frac{(1-\e^{-\frac{g \ell}{\sqrt{\pi}} })(1+ \e^{-\frac{g(L-\ell)}{\sqrt{\pi}} })}
{1+\e^{-\frac{g L}{\sqrt{\pi}} }}.
\label{eq:interval-massless}
\eeq
For $m/g= 0.2$ (right panel), 
the curves 
represent $V_f^{(0)}+V_f^{(1)}$, 
where $V_f^{(1)}$ is the $\mathcal{O}(m)$ mass perturbation given
in~(\ref{eq:potential_m1}).
The simulation results agree rather well with the analytical predictions with the same volumes.
We cannot see a clear plateau 
although the results for a larger volume ($N=21$) have a more slowly varying region
in which the values of the potential are close to that of the plateau for a very large volume ($N=101{, Lg=40}$)\footnote{{With $Lg=12$ ($N=31$), the function~$V_f^{(0)}+V_f^{(1)}$ exhibits a %clear
plateau of the  height within 3.5 percent of the height for $Lg=40$ ($N=101$).}}.
The value of the plateau for a large volume depends on $m$ and this is a result of nontrivial dynamics.
The absence of a clear 
plateau in the simulation is reasonable 
because the simulation setup
does
not have a 
parameter region that satisfies
the condition $1\ll g\ell \ll gL$ 
necessary to have a plateau.
The small discrepancy between the simulation data and the corresponding curves indicates a correction due to finite $a$ and/or systematic errors.

\begin{figure*}[t]
\centering
\includegraphics[scale=.75]{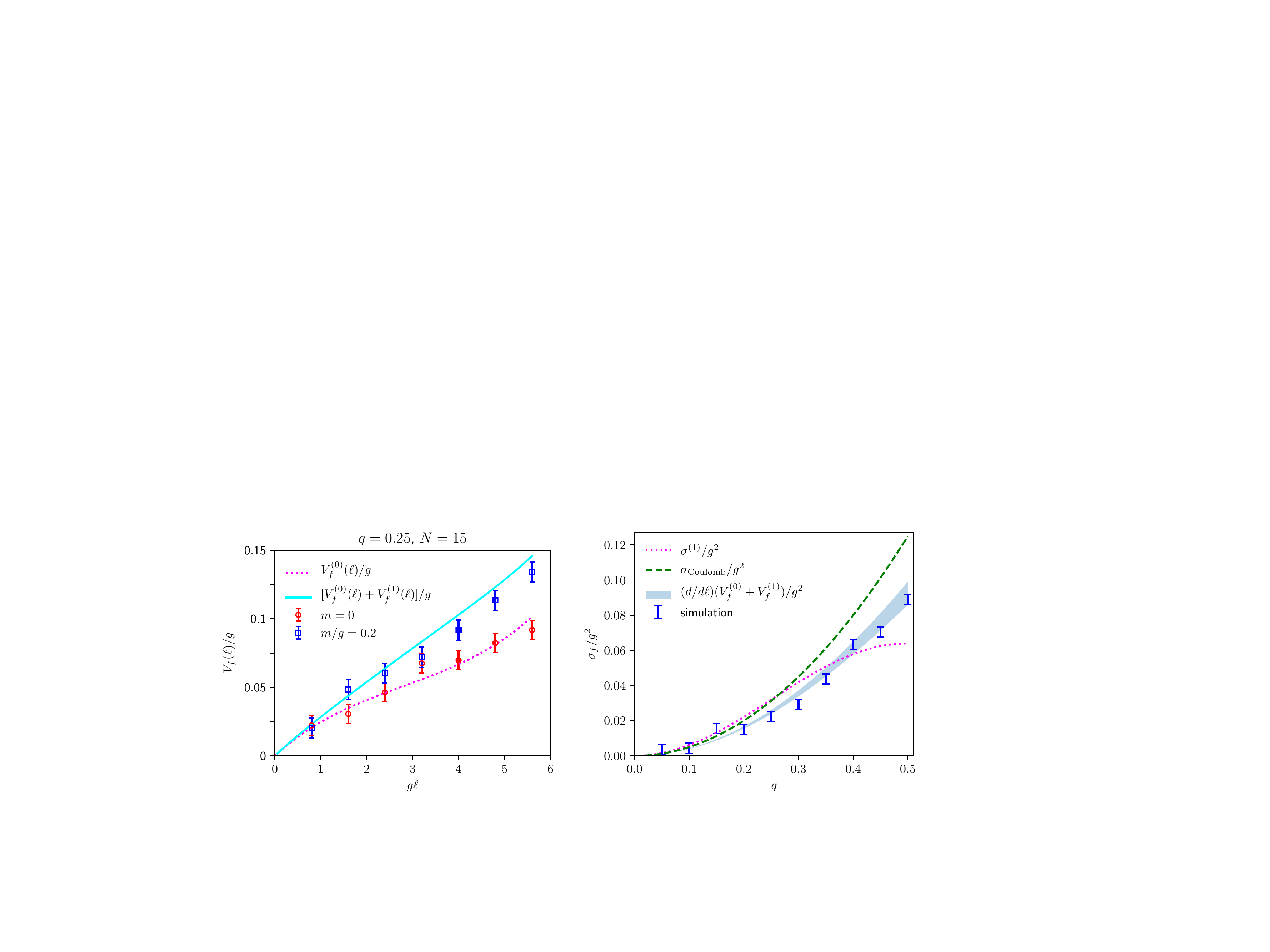}
\vspace{-5mm}
\caption{(Left): 
Results for~$V_f(\ell)$
with $n_\text{shots}=4\times 10^5$.
The error bars represent the statistical uncertainties.
The solid (cyan) and dotted (magenta) curves 
represent $V_f^{(0)}$ and $V_f^{(0)} + V_f^{(1)}$, respectively.
(Right) 
The finite-volume
analog $\sigma_f$ of the string tension for 
$m/g=0.2$.
The error bars represent the standard uncertainties from the fit.
The dashed (green) and dotted (magenta) curves
represent 
$\sigma_{\rm Coulomb} /g^2$ and $\sigma^{(1)}/g^2 $,
respectively.
The band region represents the finite-volume counterpart of $\sigma^{(1)}/g^2 $.
} 
\label{fig:tension}
\end{figure*}

In the left panel of FIG~\ref{fig:tension}, 
we depict the simulation results for $q=0.25$ as a representative of non-integer charge where the confinement (screening) behavior is expected for massive (massless) case.
We first observe
a qualitative difference between the simulation data for the massive and massless cases:
the $m/g=0.2$ case exhibits an almost linear behavior in the region $g\ell\gtrsim 3$
while the massless case does not.
This is in contrast to the integer $q$ case demonstrated in FIG.~\ref{fig:potential-intQ},
where the screening behavior is expected for any $m$.
Thus our simulation results are consistent with the above expectations
although a larger volume is needed for unambiguous demonstrations.
More quantitatively, 
we compare the simulation results with the analytical results on a finite volume.
For $m=0$ 
the simulation results
agree well
with 
the analytical prediction~$V_f^{(0)} (\ell)$ as can be seen in the left panel of FIG.~\ref{fig:potential-intQ}.
The data for $m/g=0.2$ has
moderate discrepancies from the $\mathcal{O}(m)$ analytical prediction~$V_f^{(0)}(\ell) + V_f^{(1)} (\ell)$.
In Appendix~\ref{sec:mass-deps}, 
we study the mass dependence of the potential and 
argue that their origin is likely due to the inadequacy of the $\mathcal{O}(m)$ approximation.

%%%%%%%%%%%%%%%%
%%%%%%%%%%%%%%%%

In the right panel of FIG.~\ref{fig:tension},
we plot  of the string tension a finite-$L$ ($\sigma_f$) for various values of $q$, where $\sigma_f$ is defined as the slope given by a linear fit for the energy $E(\ell)$ as a function of $\ell$.
Specifically the linear fit has been done by least squares weighted by the statistical uncertainties 
in the region 
$2.4 \leq g\ell\leq 4.8$
($\hat{\ell}=6,8,10,12$)~\footnote{%
To estimate the true string tension,
we 
should
take a larger
volume and choose a fit range away from $g\ell = 0$ and $gL$.
}.
We first observe
that our simulation data clearly deviate from the string tension~$\sigma_{\rm Coulomb}= q^2 g^2/2$
of the classical Coulomb potential
(corresponding to infinite mass),
implying that
our simulation result
exhibits confinement by non-trivial dynamics.
More quantitatively,
we again
compare the result with the $\mathcal{O}(m)$ analytical results  
both on the infinite and finite volumes.
The string tension in the infinite volume
is
\begin{align}\label{eq:sigma-1}
\sigma^{(1)}(q) 
= \frac{\e^{\gamma} m g }{2 \pi^{3/2} } 
[1-\cos (2\pi q)] ,
\end{align}
with
$\gamma$ being the Euler constant~\cite{Gross:1995bp}.
We depict
its finite-volume counterpart as the band region surrounded by the maximum and minimum of the derivative
$(d/d\ell)(V_f^{(0)}+ V_f^{(1)})/g^2$ in the range $2.4\leq g\ell\leq 4.8$
used for the linear fit~\footnote{{In Appendix~\ref{subsec:vol-dep-slope} we show that $N=35$ is enough to have good agreement between  $(d/d\ell)(V_f^{(0)}+ V_f^{(1)})/g^2$ and $\sigma^{(1)}$ for $m/g = 0.2$ and $ga = 0.4$.
%For
%With the choice of parameters 
%$m/g = 0.2$ and $ga = 0.4$, 
%we found that 
%at least with $L = 13.6 \,(N = 35)$ the slope at $\ell = L/2$ reproduces $\sigma^{(1)}$
%within $5$ percent. 
%See Appendix~\ref{subsec:vol-dep-slope} for more details.
}}.
While the simulation result is roughly consistent with
both,
it is somewhat
closer to the finite-volume result for $q\gtrsim 0.4 $.

%%%%%%%%%%%%%%%%
\begin{figure}[t]
\centering
\includegraphics[scale=0.8]{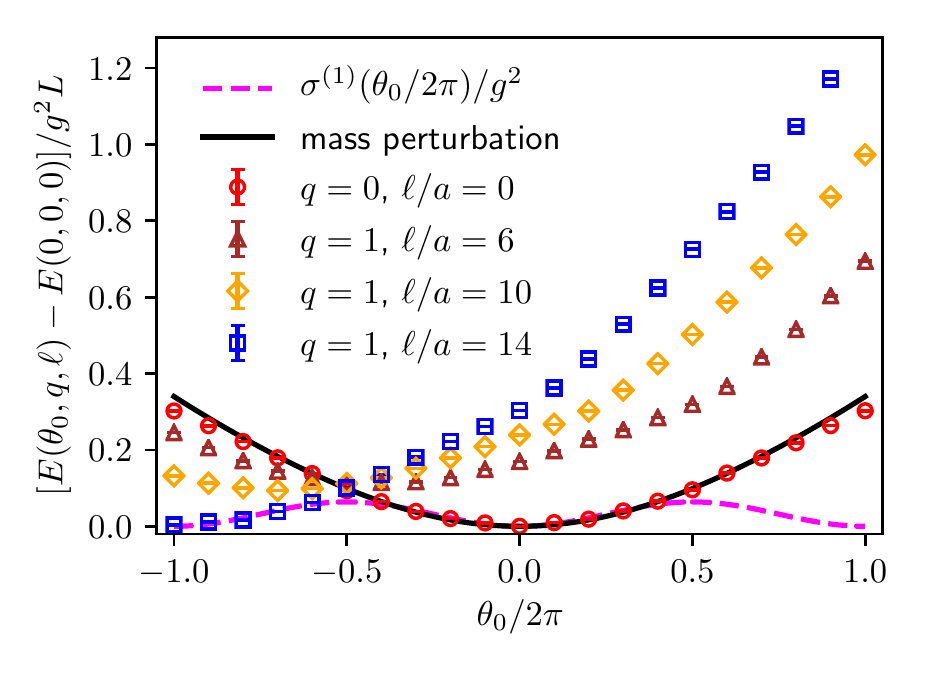}
\caption{%
Results for
$[E(\theta_0,q,\ell) - E(0,0,0)]/g^2L$ with $N=15, m/g=0.2, ag=0.4$ and $n _\text{shots}=10^5$.
The statistical uncertainties
are much smaller than the markers.
The solid (black) and dashed (magenta)
curves represent
the results of the mass perturbation theory with $q=0$ and $\sigma^{(1)} (\theta_0/2\pi ) /g^2$, respectively.
}
\label{fig:theta0-deps}
\end{figure}
%%%%%%%%%%%%%%%%

Let us turn to non-zero $\theta_0$,
which is inaccessible by
the conventional Monte Carlo approach
unless $\theta_0$ is small.
In FIG.~\ref{fig:theta0-deps}, we plot the simulation results for
$[E(\theta_0,q,\ell) - E(0,0,0)]/g^2L$ against $\theta_0$ for various values of $\ell$. 
One can read off the $\ell$-dependence of the potential from the simulation data at fixed $\theta_0$.
The simulation result
for $q=0$
agrees well with
the $\mathcal{O}(m )$ analytical result
on a finite volume
given 
explicitly
by
(\ref{eq:Ef0-def}) and (\ref{eq:energy_m1}).
For $\theta_0$ close to zero, the analytic formula $\sigma^{(1)}(\theta_0/2\pi) $
viewed as the energy density is also roughly consistent with the simulation result.

Note that the simulation results do not show the periodicity $\theta_0 \sim \theta_0 +2\pi$.
This
is expected for the open boundary condition 
because 
the theory with the theta angle $\theta_0 +2\pi k\ (k\in \mathbb{Z})$ is equivalent to 
the one with $\theta_0$ in the presence of extra probe charges $\pm k$ on the boundaries. 
Indeed in FIG.~\ref{fig:theta0-deps}, 
the blue squares 
in the range $-1\leq\theta_0/2\pi\leq 0$ 
coincide with the red circles
in the range $0\leq\theta_0/2\pi\leq 1$ up to a horizontal shift.

%%%%%%%%%%%%%%%%%%%%%%%
%%%%%%%%%%%%%%%%%%%%%%%
%%%%%%%%%%%%%%%%%%%%%%%
\section{Discussion}
\label{sec:summary}
%%%%%%%%%%%%%%%%%%%%%%%
%%%%%%%%%%%%%%%%%%%%%%%
%%%%%%%%%%%%%%%%%%%%%%%

We 
used the
classical simulator to test and demonstrate our protocol.
Let us estimate the resources needed to implement the algorithm on real quantum devices expected to be available in the near and far futures.
(For more details, see
Appendix~\ref{sec:resource}.)
In the near future,
only NISQ devices will be available.
On such devices, 
single-qubit gates will be implemented more accurately and with higher speed than two-qubit gates, 
which will be the dominant cause of the quantum error.
In our algorithm the only two-qubit gates are the CNOT gates, 
and we need $\mathcal{O}(MN^{2})$ of them to perform the discrete time evolution of $M$ Suzuki-Trotter steps in~\eqref{eq:SK_timeDep} for $N$ lattice sites. 
Thus our simulation with $N=15$ and $M=330$ would require 
roughly
a hundred thousand CNOT gates.
{This makes the NISQ implementation of our simulation implausible.}

In the far future, quantum devices 
with 
many qubits and 
less
noise
will be able to implement quantum error correction. 
In most approaches to fault-tolerance~\cite{Nielsen-Chuang}, the $T$-gate  will consume the most computational cost
among the Clifford+$T$ universal gate set.
Our algorithm demands $\mathcal{O}\left(N^{2}\log(N^{2}/\delta)\right)$ $T$-gates to implement one 
Suzuki-Trotter step 
within an accuracy $\delta$.

We have introduced two probe charges using a position-dependent theta angle.
In an Abelian gauge theory,
this method
is equivalent to the one based on a change of the Gauss law~\cite{Buyens:2015tea}.
In non-Abelian gauge theories, an approach similar to ours is to introduce probes
as
heavy fermions and measure the energy
\cite{Kogut:1981ny,Pearson:1981nz}.
In higher dimensions the quark-antiquark potential can also be estimated by Wilson loops along spatial directions as in the conventional lattice QCD.
The approach based on the inclusion of probes as part of the Hamiltonian 
is
in contrast with the conventional method based on Wilson loops,
where one suffers from a large signal-to-noise ratio in extracting the string tension, although
several smearing techniques have been applied to reduce statistical uncertainties.
It would be important to 
implement this approach in quantum simulation.

%%%%%%%%%%%%%%%%%%%%%%%
%%%%%%%%%%%%%%%%%%%%%%%
\begin{acknowledgments}
%%%%%%%%%%%%%%%%%%%%%%%
%%%%%%%%%%%%%%%%%%%%%%%
We would like to thank Mitsuhiro Kato and Yuya Tanizaki for useful conversations.
{We also thank Y.~Tanizaki for pointing out a mistake in an analytic computation in an earlier version.}
M.~H. is partially supported by MEXT Q-LEAP.
The work of E.~I. is supported by JSPS KAKENHI with Grant Numbers 
19K03875 and JP18H05407, and by the HPCI-JHPCN System Research Project (Project
ID: jh210016).
Y.~K. is supported by the U.S. Department of Energy, Office of Science, National Quantum Information Science Research Centers, Co-design Center for Quantum Advantage, under the contract
DE-SC0012704.
Discussions during the Yukawa Institute for Theoretical Physics (YITP) workshop YITP-W-20-17 on ``Quantum computing for quantum field theories" were useful for completing this work.
\end{acknowledgments}
%%%%%%%%%%%%%%%%%%%%%%%
%%%%%%%%%%%%%%%%%%%%%%%
%%%%%%%%%%%%%%%%%%%%%%%
\appendix

%%%%%%%%%%%%%%%%%%%%%%%
%%%%%%%%%%%%%%%%%%%%%%%
%%%%%%%%%%%%%%%%%%%%%%%
\section{
Systematic errors}
\label{sec:sys-error}
%%%%%%%%%%%%%%%%%%%%%%%
%%%%%%%%%%%%%%%%%%%%%%%
%%%%%%%%%%%%%%%%%%%%%%%

\begin{table*}[th]
\centering
\begin{tabular}{c|cccc}
  \hline \hline 
  $g \ell$ &  exact diag.
  &
  quant. sim.
  & systematic error %\rev{difference} 
  & statistical uncertainty
    \\
  \hline 
 0.0  &  -16.2954  &   -16.2897    &   5.7E-003  (0.03\%) &   1.00E-002   (0.06\%) \\
 0.8 &   -16.2128  &   -16.2072    &   5.6E-003  (0.03\%) &   1.00E-002  (0.06\%) \\
 1.6 &   -16.1506  &   -16.1448    &   5.8E-003  (0.04\%) & 0.99E-002 
 (0.06\%) \\
2.4 &   -16.1057  &   -16.0986
&   7.1E-003  (0.04\%) &  0.99E-002
(0.06\%) \\
 3.2 &   -16.0590  &   -16.0533    &   5.8E-003  (0.04\%) &  0.98E-002 
 (0.06\%) \\
 4.0  &  -16.0255  &   -16.0198    &   5.8E-003  (0.04\%) &  0.97E-002
 (0.06\%) \\
 4.8 &   -15.9692  &   -15.9619    &   7.3E-003  (0.05\%) &  0.97E-002
 (0.06\%) \\
 5.6 &   -15.9248  &   -15.9172    &   7.6E-003  (0.05\%) & 0.97E-002
 (0.06\%) \\
  \hline \hline
\end{tabular}
\caption{Numerical results for 
$E(\ell)/g$ obtained by exact diagonalization 
and 
quantum simulation
without statistical uncertainties for $N=15$, $ag=0.4$, $m=0$, $q=0.5$, and $\theta_0=0$.  
Systematic errors $\delta_{\rm sys}E=\delta_{\rm sys}\langle H\rangle$ and statistical uncertainties $\delta_{\rm stat}E$ for $n_{\rm shots}=10^5$ are also shown.
%\rev{with $n_{\rm shots}=10^5$  for $N=15$, $ag=0.4$, $m=0$, $q=0.5$, and $\theta_0=0$.  
%Systematic error is estimated as a difference between the exact value (second column) and the central value (third column) of the numerical data. Statistical uncertainties $\delta_{\rm stat}E$  are also shown.}
}
 \label{table:sys-error}
\end{table*}

\begin{table*}[th]
\centering
\begin{tabular}{c|cccc}
  \hline \hline 
    $g \ell$ &  exact diag.
  &  quant. sim.
  & systematic error %\rev{difference}  
  & statistical uncertainty
\\
  \hline 
 0.0 & -16.6458  & -16.6347   & 1.10E-002 (0.07\%) & 1.05E-002  (0.06\%)  \\
 0.8 & -16.5562  & -16.5757   & 1.96E-002 (0.12\%) & 1.04E-002  (0.06\%)  \\
 1.6 & -16.4770  & -16.4710   & 0.60E-002
 (0.04\%) & 1.04E-002  (0.06\%)  \\
 2.4 & -16.4052  & -16.4080   & 0.28E-002
 (0.02\%) & 1.03E-002  (0.06\%)  \\
 3.2 & -16.3327  & -16.3030   & 2.97E-002 (0.18\%) & 1.00E-002  (0.06\%)  \\
 4.0 & -16.2669  & -16.2532   & 1.37E-002 (0.08\%) & 1.00E-002  (0.06\%)  \\
 4.8 & -16.1905  & -16.1692   & 2.13E-002 (0.13\%) & 0.98E-002
 (0.06\%)  \\
 5.6 & -16.1228  & -16.0990   & 2.38E-002 (0.15\%) & 0.99E-002 
 (0.06\%)  \\
  \hline \hline
\end{tabular}
\caption{Numerical results for 
$E(\ell)/g$ obtained
for $N=15$, $ag=0.4$, $m/g=0.2$, $q=0.5$, and $\theta_0=0$.  
Statistical uncertainties for $n_{\rm shots}=10^5$ are also shown.
}
 \label{table:sys-error-2}
\end{table*}

The expectation value of an operator $\mathcal{O}$ in the true ground state~$|{\rm GS}\rangle$ is defined as
\beq
\langle \mathcal{O} \rangle = \langle {\rm GS}| \mathcal{O} | {\rm GS} \rangle.
\eeq
In our simulation
we calculate
\beq
\langle \mathcal{O} \rangle_A = \langle {\rm GS}_A | \mathcal{O} | {\rm GS}_A \rangle ,
\eeq
where $ |{\rm GS}_A \rangle$ is the approximate ground state defined in~(\ref{eq:SK_timeDep}).
Here we 
study the systematic error
\beq
\delta_{\rm sys} \langle \mathcal{O} \rangle := \langle \mathcal{O} \rangle - \langle \mathcal{O} \rangle_A,
\eeq
where $\langle \mathcal{O} \rangle$ and  $\langle \mathcal{O} \rangle_A $ 
are calculated by exact diagonalization and quantum simulation without statistical uncertainties\footnote{
This enables us to obtain the actual values of systematic errors 
rather than estimate systematic uncertainties.
}, respectively.
For exact diagonalization we 
use the
Python package
QuSpin.
For quantum simulation without statistical uncertainties we use the ``snapshot" functionality of Qiskit.

\begin{figure*}[th]
\centering\includegraphics[width=13cm]{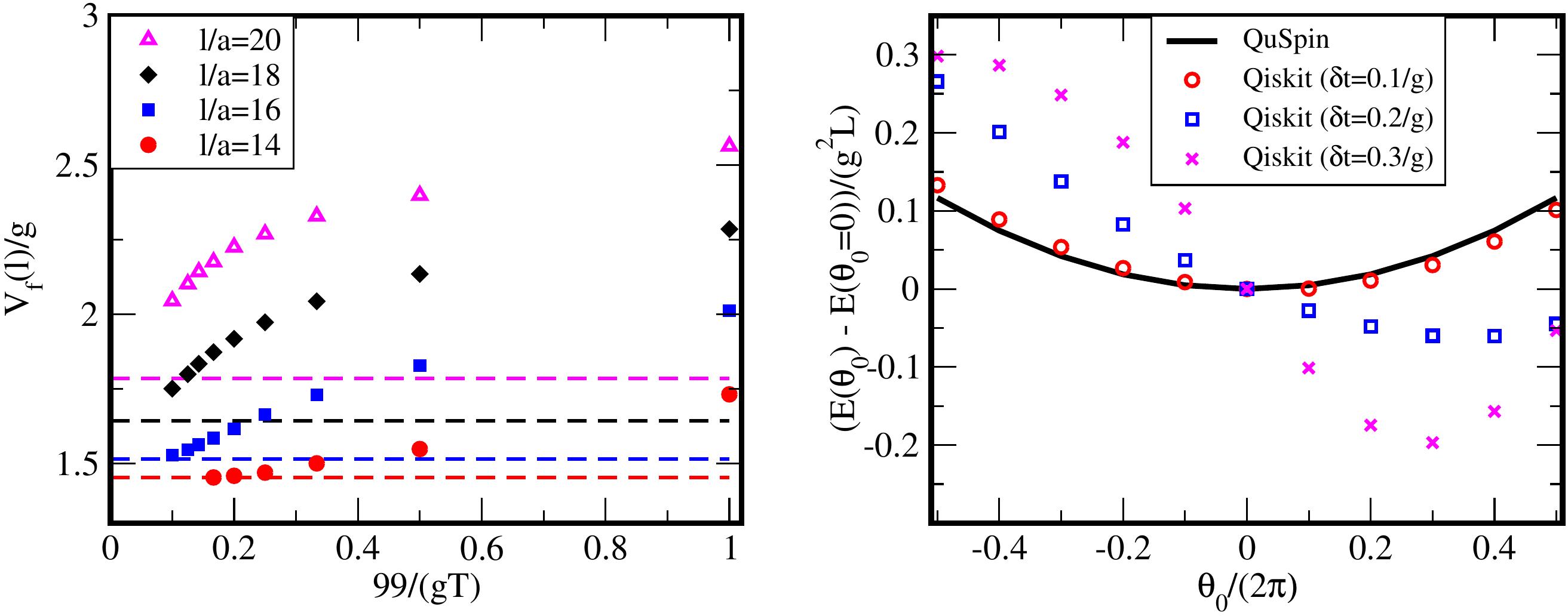}
\caption{ 
(Left): 
Dependence of the quantum simulation result for $V_f(\ell)$ on the adiabatic time~$T$ for $N=21, m/g=0.25, q=1, \theta_0=0$, and for the fixed value $\delta t =0.3 g^{-1}$.  
The dashed lines represent the results calculated by exact diagonalization.
(Right): 
Dependence of the quantum simulation 
result for $E(\theta_0)/g^2L=E(\theta_0,q=0,\ell=0)/g^2L$
on~$\theta_0$ and $\delta t$ for $ N=15$, $m/g=0.05$,
$m_0/g=0.55$, 
$T=99 g^{-1}$ 
and for a small value $a=0.1 g^{-1}$.
The exact diagonalization (QuSpin) result is represented by the solid curve.
} 
\label{fig:Tmax-error}
\end{figure*}

Besides the physical parameters of the lattice Schwinger model, adiabatic preparation requires three extra parameters: the initial mass $m_0$, the Trotter time step $\delta t$, and the adiabatic time $T$.
As in Section~\ref{sec:results}
we fix
them to $\delta t =0.3 g^{-1}$, $T=99g^{-1}$ and take $m_0 /g \approx 0.5$.\footnote{We fix $m_0 /g = 0.5$ for all the simulations in this paper except for the right panel of FIG.~\ref{fig:Tmax-error} ($m_0/g=0.55$) and the left panel of FIG.~\ref{fig:potential-mass} ($m_0/g=0.7$).}

From the numerical results shown in TABLE~\ref{table:sys-error} for $m=0$,
we see that 
the systematic
errors are smaller than the statistical uncertainties for $n_{\rm shots}=10^5$ estimated by~(\ref{def:delta-stat}).
TABLE~\ref{table:sys-error-2} shows numerical results for the particular value $m/g=0.2$ of mass.  
Systematic errors are larger than statistical uncertainties, though they are still of the same order of magnitude.
We also see the tendency that the systematic error gets larger for larger $g\ell$~\footnote{{A more detailed analysis shows that the (non-monotonic) dependence of the systematic error on $g\ell$ cannot be understood purely in terms of the adiabatic error, and that we need to take into account the Trotter error.}}.

There is a known bound on the adiabatic error (see, for example,~\cite{farhi2000quantum,Jansen2007,Wiebe_2012}).
Let us define adiabatic error $\epsilon$ as
\begin{align}
\epsilon:=\big|\big|({\bf 1}-|{\rm GS}\rangle\langle {\rm GS}|) |\widetilde{\rm GS}_A\rangle \big|\big| ,
\end{align}
where $|\widetilde{\rm GS}_A\rangle$ is the state adiabatically prepared in finite time $T$ and continuously, rather than in infinite time
as in (\ref{eq:evolution}) or discretely as in (\ref{eq:SK_timeDep}).
If the mixing between the ground state and the first excited state is the dominant source of the adiabatic error~$\epsilon$, 
then it can be bounded as
\beq
\label{eq:adiabatic-error-bound}
\epsilon \, \lesssim \, 
\max_{{t}} \frac{\lVert dH_A/dt\rVert}{\Delta^2},
\eeq
where $\Delta$ denotes the energy gap  between the ground state and the first excited state.
We note that $dH_A/dt\propto 1/T$ when $H_A$ depends on $t$ only through $t/T$. 
Generically, we expect that larger values of
($w$, $J$, $m$,
$q$,
$\theta_0$, 
$\ell$) lead to a larger value of
$||dH_{A}/dt||$, and that a larger value of $L$ and a smaller value of $a$ lead to a smaller value of $\Delta$.

%%%%%%%%%%%%%%%%
\begin{figure*}[th]
\centering
\begin{minipage}{.3\textwidth}
\includegraphics[scale=0.5]{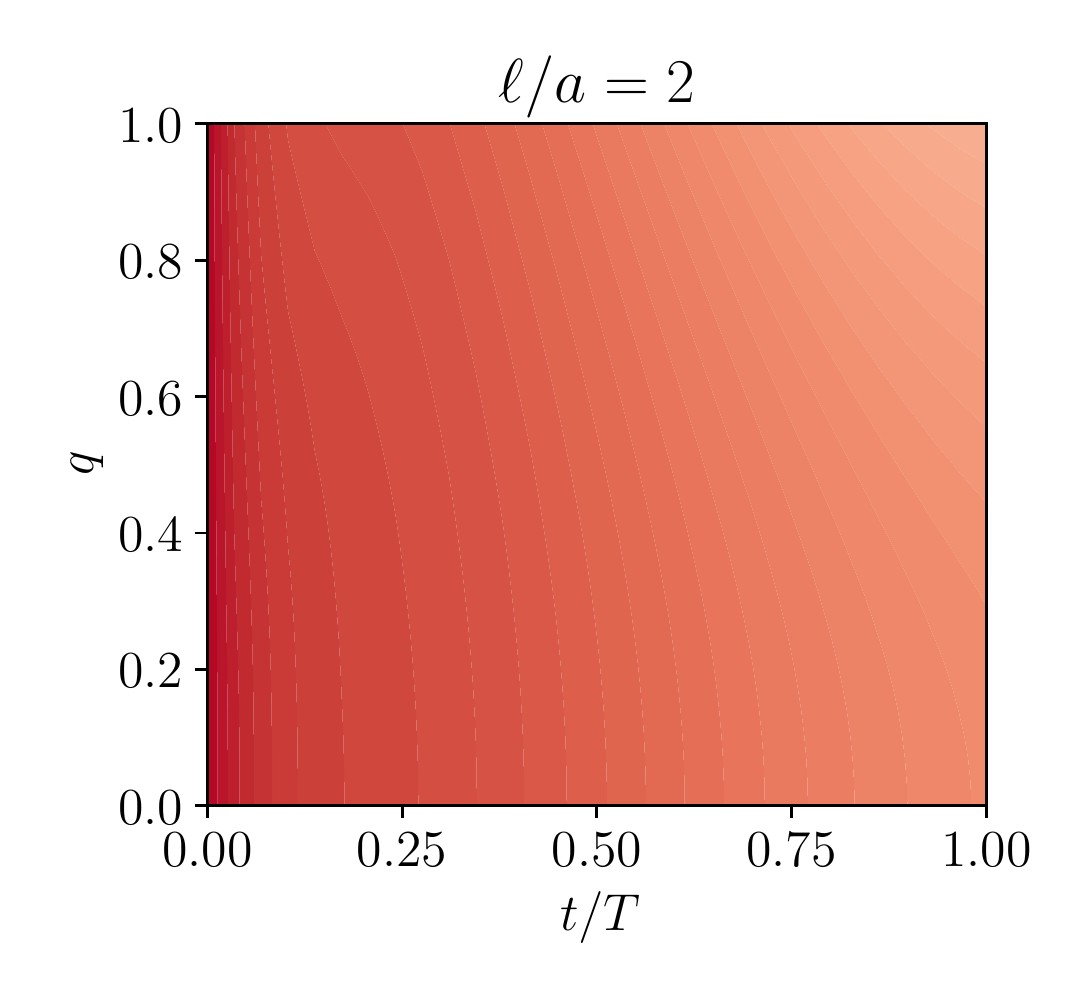}
\end{minipage}
\begin{minipage}{.3\textwidth}
\includegraphics[scale=0.5]{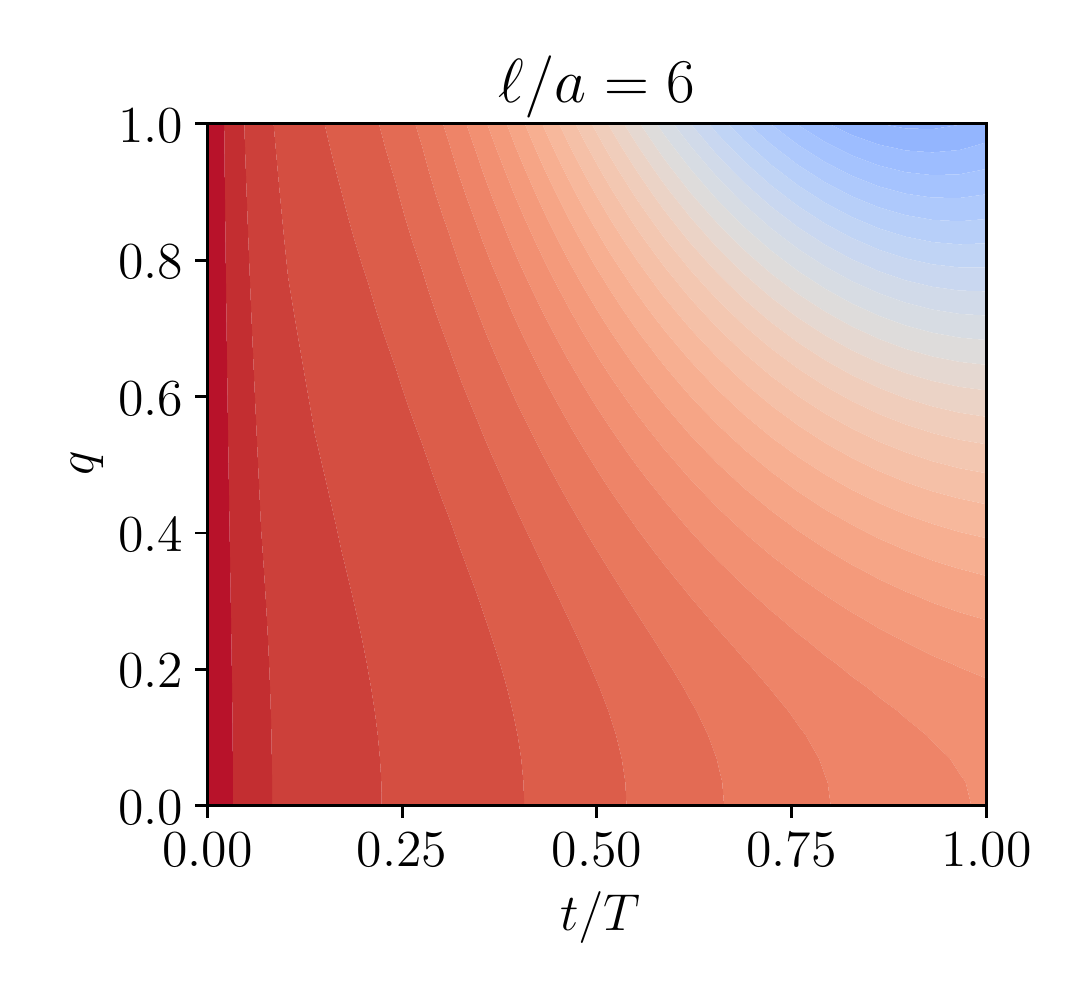}
\end{minipage}
\begin{minipage}{.35\textwidth}
\includegraphics[scale=0.5]{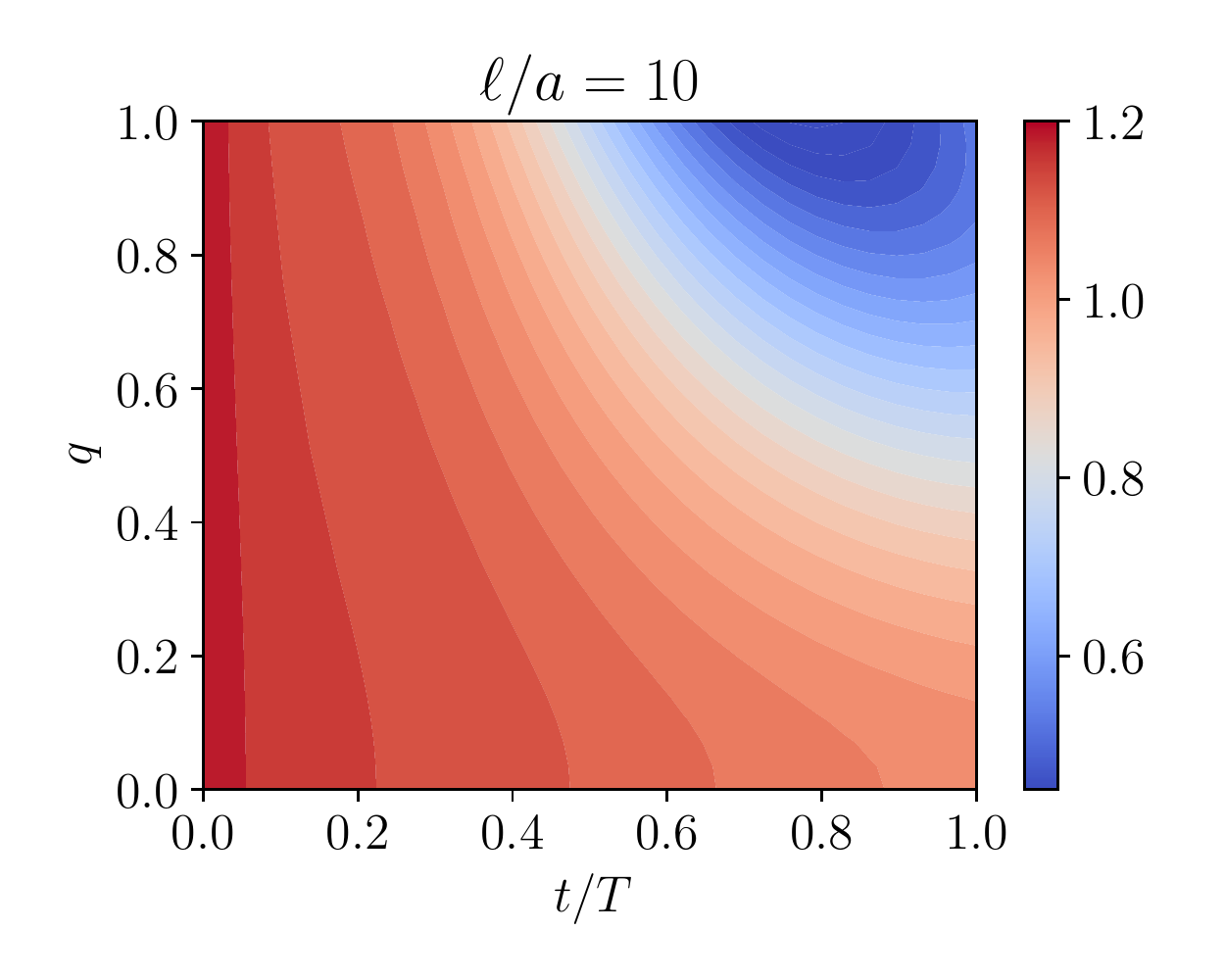}
\end{minipage}
\caption{
Density plots of
the eigenenergy of $H_A(t)/g$
for the first excited state relative to the ground state (both states in the $Q=-1$ sector)
computed for $N=15$, $m_0/g=0.5$, $m/g=0.15$, $ag=0.4$, $\theta_0=0$, and $q=1$.
Each panel shows the result for the indicated value of $\ell$.
}
\label{fig:density-plot}
\end{figure*}

%%%%%%%%%%%%%%%%

%%%%%%%%%%%%%%%%
%%%%%%%%%%%%%%%%

The density plots in FIG.~\ref{fig:density-plot} show the energy difference between
the ground state
and
the first excited state. 
The energies
are taken from the
spectrum of the adiabatic Hamiltonian $H_A (t)$, 
which  depends on $t$ through
the following $t$-dependent parameters ($0\le t\le T$): fermion mass $m_0 (1-t/T)+mt/T$, inverse lattice spacing $wt/T$, and probe charge $qt/T$.
Each horizontal path with increasing $t$ corresponds to an adiabatic variation of
the parameters~\eqref{eq:param-adiabatic} with the probe charge $q$ of the target Hamiltonian fixed. 
For the adiabatic process 
along the horizontal path passing
the region with small excitation energy (blue), a longer adiabatic time is likely to be required in order to prepare the corresponding target state with a given precision. 
In general, however, higher excited states, whose effects cannot be read off from FIG.~\ref{fig:density-plot},
also contribute to the error in the adiabatic approximation and correct the inequality~(\ref{eq:adiabatic-error-bound}).
The region with small excitation energy appears for a larger value of $q$ 
and the region tends to spread as $\ell/a$ increases. 
This suggests
that larger  $\ell$ requires longer adiabatic time $T$.

Indeed in the left panel of FIG.~\ref{fig:Tmax-error}, we find (for a fixed mass $m/g=0.25$) that a larger value of $\ell/a ( \gtrsim
14)$ 
requires a longer adiabatic time to achieve desired precision.
The quantum simulation results obtained by adiabatic preparation approach the exact diagonalization results for large $T$, linearly in~$1/T$.
This is likely due to a small energy-gap we expect for
moderately large
$g\ell$.

In the right panel of FIG.~\ref{fig:Tmax-error}, we depict the energy density as a function of $\theta_0$.
For $\delta t=0.3 g^{-1}$ and $0.2g^{-1}$ the quantum simulation results are clearly not invariant under  $\theta_0 \rightarrow -\theta_0$,
while the data obtained for $\delta t=0.1g^{-1}$ are almost invariant and reproduce the exact diagonalization results.
Exact diagonalization results shown there are for states within the same ${\rm U}(1)$ charge ($Q=\sum_n Z_n=-1$) sector as the initial state $|1010\ldots\rangle $ for adiabatic preparation. 
We confirmed that for $\theta_0/2\pi\leq -0.4$, there are states with a different value of $Q$ and with energies lower than the values shown in the figure, while for $-0.35\leq \theta_0/2\pi\leq +0.5$ there are no such states.

%\clearpage
%%%%%%%%%%%%%%%%%%%%%%%
%%%%%%%%%%%%%%%%%%%%%%%
%%%%%%%%%%%%%%%%%%%%%%%
\section{Quantum circuit}
\label{app:Q_circuit}
\subsection{Quantum operations for state preparation
}
\label{app:circuit_H}
%%%%%%%%%%%%%%%%%%%%%%%
%%%%%%%%%%%%%%%%%%%%%%%
%%%%%%%%%%%%%%%%%%%%%%%

All the quantum operations used in the main text consist of the following three elementary gates.
\begin{itemize}
\item Hadamard gate.
\begin{align}
&\begin{array}{c}
\Qcircuit @C=0.5cm @R=.3cm {
&  \gate{H} & \qw
}
\end{array}=\frac{1}{\sqrt{2}}
\begin{pmatrix}
 1 & 1 \\ 1 & -1
\end{pmatrix}
\end{align}

\item Phase (Z-rotation) gate.
\begin{align}
&\begin{array}{c}
\Qcircuit @C=0.5cm @R=.3cm {
&  \gate{R_Z(\theta)} & \qw
}
\end{array}
= \e^{-\im\frac{\theta}{2}Z}=
\begin{pmatrix}
 \e^{-\im\frac{\theta}{2}} & 0 \\ 0 & \e^{\im\frac{\theta}{2}}
\end{pmatrix}
\end{align}

\item CNOT ($CX$) gate.
\begin{align}
\begin{array}{c}
\Qcircuit @C=1em @R=.7em {
& \ctrl{1} & \qw
\\
& \targ{} & \qw
}
\end{array}
= \begin{pmatrix} 
1 & 0  & 0 & 0 \cr  
0 & 1  & 0 & 0 \cr  
0 & 0  & 0 & 1 \cr  
0 & 0  & 1 & 0 \cr  
\end{pmatrix}
\end{align}
\end{itemize}

Given a 2-qubit state on lattice sites labeled by $n\in\{0,1\}$ we decompose $\e^{-\im\alpha (X_0X_1 + Y_0Y_1)}$ and $\e^{-\im\alpha Z_0Z_1}$ in terms of the elementary gates defined above.

\begin{align}
\label{eq:XY_circuit}
&\ \e^{-\im\alpha (X_0X_1 + Y_0Y_1)} \nonumber
\\
&= 
\begin{array}{c}
\Qcircuit @C=0.2cm @R=.3cm {
&
\ctrl{1}&\gate{H}&\ctrl{1} &\gate{R_Z(2\alpha)} &\ctrl{1} &\gate{H}&\ctrl{1}&\qw
\\
&
\targ&\qw& \targ &  \gate{R_Z(-2\alpha)} & \targ & \qw &\targ &\qw
}
\end{array}
\end{align}
\begin{align}
\e^{-\im\alpha Z_0Z_1}
=
\begin{array}{c}
\Qcircuit @C=0.5cm @R=.3cm {
&\ctrl{1} &\qw&\ctrl{1} & \qw
\\
& \targ &  \gate{R_Z(2\alpha)} & \targ & \qw
}
\end{array}
\label{eq:Z_circuit}
\end{align}
Here the top and bottom lines correspond to $n=0$ and $n=1$ respectively, and $\alpha$ is a real parameter.

The initial state 
$|{\rm GS}_0\rangle=|1010\dots\rangle$ in the adiabatic time evolution can be simply constructed as
\begin{align}
|{\rm GS}_0 \rangle 
=\prod_{j=0}^{\lfloor
\frac{N-1}{2} \rfloor
} X_{2j} |00
\dots \rangle .
\end{align}
Here $\lfloor x\rfloor$ denotes the largest integer smaller than or equal to $x$.

%%%%%%%%%%%%%%%%%%%%%%%
%%%%%%%%%%%%%%%%%%%%%%%
\subsection{Measurements of $H_{XY}^{(0)}$, $H_{XY}^{(1)}$,  
$ H_{Z}$
}
\label{app:measure_H}
%%%%%%%%%%%%%%%%%%%%%%%
%%%%%%%%%%%%%%%%%%%%%%%

We spell out the measurement protocol used to compute the Hamiltonian expectation value and its statistical uncertainty for a quantum state of interest.  We consider the case where the state is a 4-qubit state for the sake of concreteness. The corresponding lattice sites are labeled by $n\in\{0,1,2,3\}$.

The term $H_{XY}^{(0)}$ consists of the operators
\begin{align}
\{X_0X_1, Y_0Y_1, X_2X_3, Y_2Y_3\}.
\end{align}
Noting that~\footnote{
We employ the following identities repeatedly:
\begin{align}
 CX_{ij}(X_iY_j)CX_{ij}=Y_iZ_j,\quad
 CX_{ij}(Y_iX_j)CX_{ij}=Y_i I_j,
 \nonumber
 \end{align}
 \begin{align}
 CX_{ij}(X_iX_j)CX_{ij}=X_i I_j,\quad
 CX_{ij}(Y_iY_j)CX_{ij}=-X_iZ_j.
 \nonumber
\end{align}
}
\begin{align}
 &H_i\, CX_{ij} (X_i X_j) CX_{ij} H_i = H_i (X_i I_j) H_i = Z_i I_j,
  \nonumber\\
 &H_i\, CX_{ij} (Y_i Y_j) CX_{ij} H_i = -H_i (X_i Z_j) H_i = -Z_i Z_j,  \nonumber
\end{align}
the expectation values that we wish to measure can be expressed as
\begin{align}
\begin{split}
 &\langle X_iX_j \rangle = \langle CX_{ij}H_i (Z_i I_j) H_i \, CX_{ij}  \rangle,
 \\
 &\langle Y_iY_j \rangle = -\langle CX_{ij}H_i (Z_i Z_j) H_i \, CX_{ij}  \rangle,
\end{split}
\end{align}
where $I_i$ stands for the identity operator acting on $i$-th site.
These quantities can be read off by the
following 
circuit.
\begin{align}
 \begin{array}{c}
\Qcircuit @C=0.5cm @R=.3cm {
\lstick{n\text{=0}}&\ctrl{1} &\gate{H}&\meter
\\
\lstick{n\text{=1}}& \targ &  \qw & \meter
\\
\lstick{n\text{=2}}&\ctrl{1} &\gate{H}&\meter
\\
\lstick{n\text{=3}}& \targ &  \qw & \meter
}
\end{array}
\label{fig:qc-xy0}
%  \nonumber
\end{align}
The four operations at the right end are the classical measurements in the $Z$ basis.
Having obtained the counts of the bit strings ``$b_0b_1b_2b_3$'' with $b_i\in\{0,1\}$ from the measurements, we calculate the expectation values {from these samples} as 
\begin{align}
\begin{split}
 &\langle X_0X_1 \rangle_{{\flat}} = \sum_{b_n}
 (1-2b_0)\frac{\text{counts}_{b_0b_1b_2b_3}}{n_\text{shots}},
 \\
 &\langle Y_0Y_1 \rangle_{{\flat}} = -\sum_{b_n}
 (1-2b_0)(1-2b_1)\frac{\text{counts}_{b_0b_1b_2b_3}}{n_\text{shots}},
 \\
  &\langle X_2X_3 \rangle_{{\flat}} = \sum_{b_n}
  (1-2b_2)\frac{\text{counts}_{b_0b_1b_2b_3}}{n_\text{shots}},
 \\
 &\langle Y_2Y_3 \rangle_{{\flat}} = -\sum_{b_n}
 (1-2b_2)(1-2b_3)\frac{\text{counts}_{b_0b_1b_2b_3}}{n_\text{shots}}.
\end{split}
\nonumber
\end{align}
Here $n_\text{shots}$ denotes the number of times the circuit is executed,\footnote{
Shots of microwave pulses execute quantum circuits on a real superconducting quantum computer.
} and counts$_{b_0b_1b_2b_3}$ denotes the number of times the bit string ``${b_0b_1b_2b_3}$'' is observed.
{The brakets~${\langle\bullet\rangle}_{{\flat}}$ stand for the expectation values from samples measured by the circuits~\eqref{fig:qc-xy0}.}

The term $H_{XY}^{(1)}$ consists of the operators 
\begin{align}
\{X_1X_2, Y_1Y_2\}.
\end{align}
Hence, the following measurement will do for the computation of their expectation values.
\begin{align}
\begin{array}{c}
% \nonumber
\Qcircuit @C=0.5cm @R=.3cm {
\lstick{n\text{=0}}& \qw &  \qw &\meter
\\
\lstick{n\text{=1}}&\ctrl{1} &\gate{H}&\meter
\\
\lstick{n\text{=2}}& \targ &  \qw & \meter
\\
\lstick{n\text{=3}}&\qw &\qw&\meter
}
\end{array}
\label{fig:qc-xy1}
\end{align}
Given the counts of the bit strings ``$b_0b_1b_2b_3$'', {we evaluate expectation values from samples as} 
\begin{align}
\begin{split}
 &\langle X_1X_2 \rangle_{{\sharp}} = \sum_{b_n}  (1-2b_1) \frac{\text{counts}_{b_0b_1b_2b_3}}{n_\text{shots}},
 \\
 &\langle Y_1Y_2 \rangle_{{\sharp}} = -\sum_{b_n}
 (1-2b_1) (1-2b_2) \frac{\text{counts}_{b_0b_1b_2b_3}}{n_\text{shots}}.
\end{split}\nonumber
\end{align}
{Here the brakets~$\langle \bullet \rangle_{{\sharp}}$ stand for the expectation values from samples measured by the circuits~\eqref{fig:qc-xy1}.}

For $H_Z$, the measurements in the computational basis allow us to compute
\begin{align}
\begin{split}
 &\langle Z_0 \rangle_{{\natural}} = \sum_{b_n}(1-2b_0)\frac{\text{counts}_{b_0b_1b_2b_3}}{n_\text{shots}},
 \\
 &\langle Z_0Z_1 \rangle_{{\natural}} = \sum_{b_n}(1-2b_0)(1-2b_1)\frac{\text{counts}_{b_0b_1b_2b_3}}{n_\text{shots}},
\end{split}\nonumber
\end{align}
{where brackets $\langle\bullet\rangle_{{\natural}}$ denote the expectation values from samples measured in the computational basis.}
Combining these results leads to the expectation value of 
the total Hamiltonian~\eqref{eq:H_decomposition0}.

We can also evaluate statistical uncertainties from the data obtained for computing the vacuum expectation value of the Hamiltonian
as follows.
First, we compute the expectation values of the squares of each term in the Hamiltonian from those data. For example, the term $(H_{XY}^{(0)})^{2}$ contains the term $X_{0}X_{1}X_{2}X_{3}$ and its expectation value is given by
\begin{align}
\langle
X_{0}X_{1}X_{2}X_{3}
\rangle_{{\flat}}
&=
\sum_{b_{n}}
(1-2b_{0})(1-2b_{2})
\notag
\\
&\qquad\qquad\times
\frac{\text{counts}_{b_{0}b_{1}b_{2}b_{3}}}{n_\text{shots}}.
\nonumber
\end{align}
Other terms in $\langle(H_{XY}^{(0)})^{2}\rangle_{{\flat}}$, $\langle(H_{XY}^{(1)})^{2}\rangle_{{\sharp}}$,
$\langle(H_{Z}^{(0)})^{2}\rangle_{{\natural}}$ can be evaluated in a similar manner.
We then obtain the statistical uncertainty $\delta_{\text{stat}}E$ from these quantities as
\begin{equation}\label{def:delta-stat}
(\delta_{\text{stat}}E)^{2}
=
\frac{(\hat{\delta}E)^{2}}{n_\text{shots}-1},
\end{equation}
where~\footnote{
{Let us consider more general cases first.
Suppose that we want to evaluate the sum of non-commutative operators $A,B$, and we measure two operators independently.
Then, the variance of $A+B$ is evaluated as
\begin{align}
&\quad\sum_{A,B} f(A)g(B) (A+B)^2 
- \left(\sum_{A} f(A)\right)^2 
\notag\\&\qquad
- 2\left(\sum_{A} f(A)\right) \left(\sum_{B} g(B)\right)
- \left(\sum_{B} g(B)\right)^2
\notag
\\
&= \sum_{A} f(A)A^2 + \sum_{B} g(B)B^2
\notag
\\&\qquad
- \left(\sum_{A} f(A)A\right)^2 - \left(\sum_{B} g(B)B\right)^2,\notag
\end{align}
where $f(A),g(B)$ are the distribution functions for $A$ and $B$.
So we do not need to consider cross terms.
In our protocol, we independently measure the expectation values of $H_{XY}^{(0)}, H_{XY}^{(1)}$ and $H_{Z}^{(0)}$, thus the statistical uncertainty of the total energy can be evaluated as \eqref{def:delta-stat}
and
\eqref{def:delta-energy}
without cross terms.
% In our protocol, we separately measure the expectation values of $H_{XY}^{(0)}, H_{XY}^{(1)}$ and $H_{Z}^{(0)}$.
% Thus we do not need to consider cross terms between them in evaluating the statistical uncertainty.
}}
\begin{align}
&(\hat{\delta}E)^{2} 
:=
\langle(H_{XY}^{(0)})^{2}\rangle_{{\flat}}+\langle(H_{XY}^{(1)})^{2}\rangle_{{\sharp}}+\langle(H_{Z}^{(0)})^{2}\rangle_{{\natural}}
\notag\\
&\qquad%\qquad
-\left(\langle H_{XY}^{(0)}\rangle_{{\flat}}\right)^{2}
-\left(\langle H_{XY}^{(1)}\rangle_{{\sharp}}\right)^{2}
-\left(\langle H_{Z}\rangle_{{\natural}}\right)^{2}.
\label{def:delta-energy}
\end{align}

%%%%%%%%%%%%%%%%%%%%%%%
%%%%%%%%%%%%%%%%%%%%%%%
\subsection{Resource estimation}
\label{sec:resource}
%%%%%%%%%%%%%%%%%%%%%%%
%%%%%%%%%%%%%%%%%%%%%%%
Let us estimate the computational resource required 
for our simulation, 
closely following~\cite{Shaw2020quantumalgorithms}.
The circuit~\eqref{eq:XY_circuit} has four CNOT gates and two $R_{Z}$-gates, while the circuit~\eqref{eq:Z_circuit} has two CNOT gates and one $R_{Z}$-gate.
Each step 
in the Suzuki-Trotter decomposition~\eqref{eq:SK_timeDep}
has 
$2(N-1)$ subcircuits of the form~\eqref{eq:XY_circuit} and 
$\frac{1}{2}(N-1)(N-2)$ subcircuits of the form~\eqref{eq:Z_circuit}, 
and $N$ $R_Z$-gates besides. 
Thus, for odd $N$,
we need $(N-1)(N+6)$ CNOT gates and $\frac{1}{2}(N^{2}+7N-6)$ $R_{Z}$-gates for each step.

As mentioned in Section~\ref{sec:summary},
 CNOT gates are the main source of errors in NISQ devices.
Thus the number $(N-1)(N+6)$ of CNOT gates
is crucial for implementation on NISQ devices.
On the other hand, 
in most approaches to fault-tolerance~\cite{Eastin2009,Bravyi2012},
the most costly components are non-Clifford operations such as $T$-gates,
where $T={\rm diag}(1,\e^{\frac{{\rm i}\pi}{4}} )$ in a matrix representation.
In our algorithm $T$-gates
appear only in implementing $R_{Z}$-gates.
They can be implemented by using the Repeat-Until-Success method~\cite{Paetznick2014,Bocharov_2015}
with $1.15\log_{2}(2/\delta)$ $T$-gates within an accuracy $\delta$:
\begin{equation}
\|
R_{Z}-\tilde{R}_{Z}
\|<\delta,
\end{equation}
where $\tilde{R}_{Z}$ approximates $R_{Z}$-gates in a fault-tolerant manner and $\|\cdot\|$ stands for the spectral norm.
Thus we need $0.575(N^{2}+7N-6)
\log_2\left((N^{2}+7N-6)/\delta\right)$ $T$-gates to implement each step
in the Suzuki-Trotter decomposition~\eqref{eq:SK_timeDep}.

In our simulations, the computation of the potential $V_f$ within the $\mathcal{O}(10\%)$ accuracy requires $n_\text{shots}=4\times 10^5$ measurements for a non-integer probe charge.
In our quantum simulation on a classical simulator, it takes about $3$ hours for $N=21$ to obtain the potential in the range $0 \le  \ell /a \le 20$ ($11$ data points) with $n_\text{shots}=10^5$
on a typical personal computer (such as iMac).

%%%%%%%%%%%%%%%%%%%%%%%
%%%%%%%%%%%%%%%%%%%%%%%
%%%%%%%%%%%%%%%%%%%%%%%
\section{Comments on symmetries}
\label{app:parity}
%%%%%%%%%%%%%%%%%%%%%%%
%%%%%%%%%%%%%%%%%%%%%%%
%%%%%%%%%%%%%%%%%%%%%%%
We comment on discrete symmetries in the lattice theory.
In the continuum theory, the net effect of the parity (or charge conjugation) transformation is to flip the sign of~$\theta_0$ because it acts as $F_{01} \rightarrow - F_{01}$.
Therefore, the continuum theory in the absence of probes is manifestly parity invariant at $\theta_0=0$.
On a space without boundary, the theory at $\theta_0=\pi$ also has parity invariance because of the periodicity $\theta_0 \sim \theta_0 +2\pi$
while in the case with boundaries,
the theories at $\theta_0=\pi$ and $-\pi$ differ by boundary terms in the action.

The situation in the lattice theory is more intricate.
An analogue of the parity transformation on the lattice is
\begin{align}
&
\chi_n \rightarrow i(-1)^n \chi_{N-1-n} ,\quad 
U_n \rightarrow U^{-1}_{N-2-n},
\nonumber\\
&
\qquad\qquad\qquad
L_n \rightarrow -L_{N-2-n} .
\label{eq:latticeP}
\end{align}
For the periodic boundary condition, in which $N$ is always even,
the periodic analog of the lattice Hamiltonian~(\ref{eq:fermion-Hamiltonian})
is invariant under this transformation for $\theta_0 =0$ and $\pi$.\footnote{
The $(-1)^n/2$ term 
in the Gauss law constraint~(\ref{eq:Gauss_lattice}) 
violates the invariance under (\ref{eq:latticeP}) for even $N$.
}
With the open boundary condition, the situation is different for 
odd $N$ and even $N$.
For odd $N$, the net effect of the transformation \eqref{eq:latticeP} is to flip the sign of the theta angle as in the case of the continuum theory 
: it is parity invariant for $\theta_0=0$, and the cases with $\theta_0=\pi$ and $-\pi$ 
differ only by boundary terms.
However, for even~$N$,
the transformation \eqref{eq:latticeP} flips not only the theta angle but also the mass $m$.
This implies that the lattice theory with even $N$ and non-zero mass
does not have parity symmetry for any value of $\theta_0$.
Thus we take $N$ to be odd in this work.
In the presence of symmetrically located probe charges 
in (\ref{eq:theta_config}), 
the net effect of the parity transformation \eqref{eq:latticeP} is to flip the signs of both $\theta_0$ and $q$.

With the choice of the decomposition~(\ref{eq:H_decomposition0}),
the time evolution
in~(\ref{eq:SK_timeDep})
respects the ${\rm U}(1)$ symmetry generated by $Q=\sum_{n=0}^{N-1}Z_n$,\footnote{See~\cite{Yamamoto:2021vxp} for another decomposition that preserves  ${\rm U}(1)$.} while it violates the parity symmetry,
which is only approximately restored for small $\delta t$.
Whether it is ${\rm U}(1)$ or parity, a symmetry can be a useful property for simulation~(see {\it e.g.} \cite{Tran_2021}).
In general an adiabatic process prepares a state within the same charge sector as the initial state.
We checked by exact diagonalization that the lowest-energy states among all charge sectors do have $Q=-1$
for the parameters discussed in FIGs.~\ref{fig:potential-intQ} and~\ref{fig:tension}, and in Tables~\ref{table:sys-error} and \ref{table:sys-error-2} in Appendix~\ref{sec:sys-error}.
But for some of the parameters considered in FIGs.~\ref{fig:theta0-deps} and~\ref{fig:Tmax-error}, we found that the lowest-energy states have different values of $Q$.

%%%%%%%%%%%%%%%%%%%%%%%%%%%%%%%
%%%%%%%%%%%%%%%%%%%%%%%%%%%%%%%
%%%%%%%%%%%%%%%%%%%%%%%%%%%%%%%
\section{Continuum Schwinger model}
\label{sec:continuum}
\subsection{Summary of results}
\label{app:summary}
%%%%%%%%%%%%%%%%%%%%%%%%%%%%%%%
%%%%%%%%%%%%%%%%%%%%%%%%%%%%%%%
%%%%%%%%%%%%%%%%%%%%%%%%%%%%%%%
We summarize some analytic results for two probe charges $\pm q$ separated by distance $\ell$ in the continuum Schwinger model.
Some of the results are new.

We begin with the theory in the infinite-volume limit.
In this case, there are some results by mass perturbation theory in the literature \cite{Iso:1988zi,Gross:1995bp,Adam:1997wt}.
The potential in the massless ($m=0$) theory is given by~\cite{Iso:1988zi} 
\beq
V^{(0)}(\ell) = \frac{q^2 g^2}{2 \mu} (1-\e^{-\mu \ell}),
\label{eq:Gross-massless-appendix}
\eeq
where $\mu = g/\sqrt{\pi}$.
The~$\mathcal{O}(m)$ correction to the potential for $\ell\gg 1/\mu$ is given by
\begin{align}
\hspace{-2mm}
V^{(1)}(\ell) 
= m \Sigma (1-\cos (2\pi q)) \ell +c_q m +o\left( 1/\ell \right) ,
\label{eq:Gross-massive}
\end{align}
where $\Sigma=e^\gamma g/(2\pi^{3/2})$.
The first term giving the string tension has been known in the literature (see {\it e.g.}
\cite{Gross:1995bp})
while the second term has not.
In Appendix~\ref{app:bosonization}, we find that the constant~$c_q$ is given by 
\begin{align}
 c_q &= \frac{2\e^\gamma}{\pi}\cos(\theta_0+\pi q)\big[
    \cos(\pi q) {\rm Cin}(\pi q)
    \nonumber\\
    &\qquad\qquad\qquad\qquad\qquad
    -\sin(\pi q){\rm Si}(\pi q)
    \big],\label{eq:cq}
\end{align}
where ${\rm Cin}(z):= \int_0^z ds [1-\cos(s)]/s$ and ${\rm Si}(z):= \int_0^z ds \sin(s)/s$ are known as the cosine and sine integral functions, 
respectively.\footnote{
The function~${\rm Cin}(z)$ is related to a more common function~${\rm Ci}(z):= -\int_z^\infty ds \cos(s)/s$ as ${\rm Cin}(z)=\gamma+\log z-{\rm Ci}(z)$.}
Note that $c_q$ is non-zero even when $q$ is an integer.
This implies that 
the potential for large $\ell$ in integer $q$ case approaches 
a non-zero constant which is given by $ c_q m$ up to an $\mathcal{O}(m^2 )$ correction.
Therefore the value of the screening potential on the plateau in the massive case 
differs from the one in the massless case, 
as can be confirmed for the analytic curves on the right panel of FIG.~\ref{fig:potential-intQ}.%
\footnote{
For $m/g=0.2$, $c_{q=1} m /g\simeq 0.37$, which is the difference between the limiting values (for large $g\ell$) of the dashed curves in green and purple in
FIG.~\ref{fig:potential-intQ}.
}
The~$\mathcal{O}(m^2)$ correction for $\ell\gg 1/\mu$ is~\cite{Adam:1997wt}
\begin{align}
V^{(2)}(\ell) &= m^2\Big[ \pi   \left( \frac{\Sigma}{2g^2} \right)^2 
%\hspace{-1mm}
\mu_0^2 \mathcal{E}_+ (1-\cos(4 \pi q))  g \ell \nonumber\\
&\qquad\qquad\qquad\qquad
+ \mathcal{O}(1)\Big] ,
\label{eq:Adam}
\end{align}
where $\mu_0^{2} \mathcal{E}_+ \simeq -8.9139$.%
\footnote{
The precise definition is given by Eq.(64) in \cite{Adam:1997wt} as
$\mu_0^{2} \mathcal{E}_+ =2\pi \int_0^\infty dr r(\e^{-2K_0 (r)}-1)$
with the modified Bessel function $K_0 (r)$ of the second kind.
}
This formula is consistent with the expectation that 
the string tension vanishes exactly when $q$ is an integer~\cite{Coleman:1975pw}.

Next let us consider the  Schwinger model on the finite interval $[0,L]$.
As explained in Appendix~\ref{app:bosonization},
we take an appropriate boundary condition, 
which corresponds to the continuum limit of the lattice model studied in the main text.
For this boundary condition, we derive the following results for the probe charges $\pm q$ placed at $x=(L\mp \ell)/2$.
For $m=0$, the ground state energy is
\begin{equation} \label{eq:Ef0-def}
E_f^{(0)}(\theta_0,q,\ell):=
\sum_{n=1}^\infty  \frac{L\mu^2}{16\pi} \frac{k_n^2}{\omega_n^2}(\Theta_n^{(q)})^2,
\end{equation}
where
\begin{align}
\Theta_{n}^{(q)}& =
\frac{1-(-1)^n}{n}\left( \frac{2 \theta_0}{\pi } + (-1)^{\frac{n-1}{2}} 4q \sin \left( \frac{\pi  n \ell}{2L}  \right)\right)
\nonumber\\
&\qquad
+\frac{1+(-1)^n}{n}.
\label{Theta-q-n}
\end{align}
For $\theta_0=0$, the potential $V_f^{(0)}(\ell) :=E_f^{(0)}(0,q,\ell)-E_f^{(0)}(0,q,0) $ simplifies to
\beq
V_f^{(0)}(\ell) = \frac{q^2 g^2}{2 \mu} \frac{(1-\e^{-\mu \ell})(1+ \e^{-\mu (L-\ell)})}{1+\e^{-\mu L}}.
\label{eq:interval-massless-appendix}
\eeq
One can easily see that this reproduces \eqref{eq:Gross-massless-appendix}
in the infinite-volume limit $L\rightarrow\infty$ with $1\ll \mu\ell \ll \mu L$.
The $\mathcal{O}(m)$ correction to the energy is given by
\begin{align}
&E_f^{(1)}(\theta_0,q,\ell) = -m \Sigma
\int_0^L 
dx \
\lambda(x)
\nonumber\\
&\qquad
\times
 \cos \left[{2\sqrt{\pi}\phi_0-} \sum_{n=1}^\infty \frac{\mu^2\Theta^{(q)}_n}{\mu^2 + k^2_n}  \sin(k_n x) \right],
\label{eq:energy_m1}
\end{align}
where $k_n=\pi n/L$,
\begin{align}
\lambda(x) &:= 
\lim_{\Lambda\rightarrow \infty} 
\exp\Bigg[  \sinh^{-1}
\left(
\frac{\Lambda}{\mu}\right) 
\nonumber\\
&\qquad\qquad
-\sum_{n=1}^{\lfloor L\Lambda/\pi\rfloor} \frac{2\pi}{L} \frac{\sin^2(k_nx)}{\sqrt{\mu^2+ k_n^2}}
\Bigg],\label{def:lambda}
\end{align}
 $\Lambda$ is a UV cutoff sent to infinity, and $\lfloor x\rfloor$ denotes the largest integer smaller than or equal to $x$.
The $\mathcal{O}(m)$ correction to the potential for $\theta_0=0$ is by definition
\begin{align}
V^{(1)}_f(\ell) &:= 
E_f^{(1)}(0,q,\ell) - E_f^{(1)}(0,q,0).
\label{eq:potential_m1}
\end{align}

We also show in Appendix~\ref{app:bosonization} that for $m/g\gg 1$ and for $\theta_0=0$, the potential~$ V_f(\ell)$ becomes
\begin{equation}\label{eq:large-mass-potential}
    \frac{q^2g^2}{2}\left[1- \frac{\e^{-\gamma } g}{2 \sqrt{\pi } m} + \mathcal{O}((g/m)^2)\right]\ell .
\end{equation}
The leading term is the Coulomb potential of the pure ${\rm U}(1)$ gauge theory as expected.

%%%%%%%%%%%%%%%%%%%%%%%%%%%%%%%
%%%%%%%%%%%%%%%%%%%%%%%%%%%%%%%
\subsection{Bosonized Schwinger model on an interval}
\label{app:bosonization}
%%%%%%%%%%%%%%%%%%%%%%%%%%%%%%%
%%%%%%%%%%%%%%%%%%%%%%%%%%%%%%%
Here we study the bosonized Schwinger model on an
interval and derive some of the analytic results used in the main text and summarized Appendix~\ref{app:summary}. 
Let us consider the Schwinger model on the spacetime $\mathbb{R}\times[0,L]$.
We parametrize the time by $t= x^0$ and the interval~$[0,L]$ by~$x= x^1$.
After bosonization the theory with a position-dependent theta angle~$\Theta(x)$ is described by the Lagrangian density
\begin{align} \label{eq:Lag-bosonized}
\mathcal{L}_b&=-\frac14 F_{\mu\nu}F^{\mu\nu} 
+\frac{g}{4\pi}\Theta(x) \epsilon^{\mu\nu}F_{\mu\nu} 
+ \frac{g}{\sqrt\pi} \epsilon^{\mu\nu} A_\mu \partial_\nu \phi 
\nonumber\\
&\quad + \frac12 \partial_\mu\phi\partial^\mu\phi 
+ m g \frac{e^\gamma}{2\pi^{3/2}}  \cos(2\sqrt\pi\phi) .
\end{align}

We first specify the boundary conditions for the fields.
For $\phi$, we impose
\begin{equation}
\left. \phi \right|_{x=0}  =\sqrt\pi w_0 ,\quad
\left. \phi \right|_{x=L} =\sqrt\pi w_1 ,  
\end{equation}
where $w_0$ and $w_1$ are real (not necessarily integral) constants 
specifying the boundary conditions.
In the main text we set these constants to the specific values $w_0=1/4$ and $w_1=-1/4$ for the reasons we explain below~\eqref{Theta-n-appendix}.
Regarding the gauge field, it must be consistent with the Gauss law constraint~$\delta (\int d^2x\mathcal{L}_b)/\delta A_0=0$,
which reads
\begin{equation}\label{boson-Gauss}
\partial_1\left(F_{01}+ \frac{g}{2\pi} (\Theta +2\sqrt\pi \phi) \right)=0 \,.
\end{equation}
This is solved by
\begin{equation}
F_{01} = \frac{g}{2\pi} (\Theta +2\sqrt\pi \phi)
  + f(x^0) ,
\end{equation}
where $f(x^0)$ is $x$-independent.
As in Section~\ref{sec:qubit-Schwinger}
we work in the temporal gauge $A_0=0$.
We choose the boundary condition on $A_1$ such that $f(x^0) = 0$.

Denoting the canonical momentum conjugate to $\phi$ by~$\Pi_\phi$,
we get the Hamiltonian density
\begin{align}\label{eq:Ham-density-bosonized}
\mathcal{H}_b(x) &= \frac12\Pi_\phi^2+\frac12(\partial_x\phi)^2 
+\frac{g^2}{2\pi} \left(\phi + \frac{\Theta(x)}{2\sqrt\pi}\right)^2
\nonumber\\
&\ \qquad
- mg \frac{\e^\gamma}{2\pi^{3/2}} \cos(2\sqrt\pi \phi)
\end{align}
and the Hamiltonian $H_b =\int_0^L dx \,\mathcal{H}_b(x) $.
We note that the numerical coefficient~$\e^\gamma/2\pi^{3/2}$ for the cosine terms in~(\ref{eq:Lag-bosonized}) and~(\ref{eq:Ham-density-bosonized}) is appropriate only if ``$\cos(2\sqrt\pi \phi)$'' in~(\ref{eq:Ham-density-bosonized}) is interpreted as normal-ordered with mass $\mu=g/\sqrt{\pi}$ on an infinite spatial line.
This will be important below.

We now set $m=0$ to compute the energy by the mass perturbation theory.
Let us define $\phi_0(x)$ and $\hat\phi(x)$ by
\begin{align}
\phi_0(x)&= \sqrt\pi w_0+\sqrt\pi (w_1-w_0)\frac{x}{L},
\nonumber \\
  \hat  \phi(x)&=\phi(x)-\phi_0(x).
  \nonumber
\end{align}
We also define
\begin{equation}
    \hat\Theta(x)\equiv \Theta(x)+2\sqrt\pi \phi_0(x).
\end{equation}
We expand the operators in terms of the Fourier coefficients~$\phi_n, \Pi_n, \Theta_n$ as
\begin{align}\label{eq:Fourier-coeff}
&\hat\phi(x)  = \sum_{n=1}^\infty \phi_n \sin(k_n x),
\quad
\Pi_\phi (x) = \sum_{n=1}^\infty \Pi_n \sin(k_n x),
\nonumber\\
& \qquad\qquad\qquad
 \hat\Theta(x) = \sum_{n=1}^\infty \Theta_n \sin(k_n x) .
\end{align}
Here we extended~$\hat\Theta(x)$ to a (possibly discontinuous) odd periodic function with period $2L$.
We have the canonical commutation relations
$[\phi_n , \Pi_{n'}] =  ({\rm 2i}/L) \delta_{nn'} $.
The Hamiltonian can be written as
\begin{align}
H_b&=\frac{\pi(w_1-w_0)^2}{2L}
\nonumber\\
&\
+\sum_{n=1}^\infty 
\Big[
\omega_n \left(a_n^\dagger a_n
+\frac12\right) 
+
\frac{L\mu^2}{16}
\frac{k_n^2}{\omega_n^2}\Theta_n^2
\Big],
\end{align}
where $\omega_n =  \sqrt{
\mu^2 + k_n^2
}
$ and
\begin{equation} \label{interval-an}
a_n = \frac{\sqrt{L\omega_n}}{2}\left(\phi_n + \frac{\mu^2 \Theta _n}{2 \sqrt \pi \omega_n^2}\right)+ \frac{\rm i}{2} \sqrt{\frac{L}{\omega_n}} \Pi_n   \,.
\end{equation}
We have $[a_n,a_{n'}^\dagger ] = \delta_{nn'}$.
The ground state~$|0\rangle$ satisfies $a_n|0\rangle=0$, and its energy is 
\begin{align}
\langle 0 | H_b |0\rangle &=
\frac{\pi(w_1-w_0)^2}{2L}
\nonumber\\
&\quad
+
\sum_{n=1}^\infty \left( \frac{\omega_n}2 + \frac{L\mu^2}{16\pi} \frac{k_n^2}{\omega_n^2}\Theta_n^2 \right)
.\label{eq:ground-state-energy}
\end{align}
The sum of the first term in the bracket is the ($L$-dependent) Casimir energy.
It is UV-divergent but independent of $\Theta$.
Therefore we focus on the sum of the second term.
We take
\begin{equation}\label{eq:Theta-x}
\Theta(x) =\left \{
\begin{array}{cc}
\displaystyle  \theta_0 + 2\pi q & \text{for }   \frac{L-\ell}{2}\leq x\leq  \frac{L+\ell}{2} ,\\
\displaystyle \theta_0 & \text{otherwise.}
\end{array}
\right.
\end{equation}
The Fourier coefficients for $\hat\Theta$ are then
\begin{align}
\Theta_{n} &=\frac{1-(-1)^n}{n}\left[ \frac{2 \theta_0}{\pi } + (-1)^{\frac{n-1}{2}} 4q \sin \left( \frac{\pi  n\ell}{2L}  \right)\right]
\nonumber\\
&\qquad
+4\frac{w_0 - (-1)^n w_1}{n}.
\label{Theta-n-appendix}
\end{align}

We claim that the lattice Schwinger model described in
Section~\ref{sec:qubit-Schwinger}
corresponds, in the continuum limit, to the choice $w_0=1/4$ and $w_1=(1/2)Q +1/4$, where $Q:=\sum_{n=0}^{N-1}Z_n$.
To see this, let us compare the Gauss law constraints $L_n -L_{n-1} = [Z_n + (-1)^n]/2$ and (\ref{boson-Gauss}) in the spin and bosonized formulations, respectively.
The correspondence $
L_n \leftrightarrow  (1/g)F_{01} -(1/2\pi)\Theta$ suggests that $\phi(x)/\sqrt\pi$ is the mean field (spatial average) for $(1/2)\sum_{i=0}^n[Z_i+(-1)^i]$.
By the symmetry under $Z_i\rightarrow - Z_i$ present for $g=0$, the mean value of $Z_i$ on sites near (compared with $1/g$) the boundary must vanish.
The mean value of $(1/2)\sum_{i=0}^n(-1)^i$ is $1/4$.
This leads to our claim, which can be explicitly confirmed by comparing the results for the charge density computed by DMRG and by bosonization~\cite{Interval}.

We now restrict to odd $N$ and $Q=-1$, so that $w_1=-1/4$.
Substituting the values of $w_0$ and $w_1$ to~(\ref{Theta-n-appendix}) gives (\ref{Theta-q-n}).
The $q$-dependent part of the ground state energy~(\ref{eq:ground-state-energy}) for $\theta_0=0$
\begin{align}
 \frac{4  g^2q^2}{L} \sum_{j=0}^\infty \frac{ \sin^2 \left(k_{2j+1} \ell/2\right) }{ \omega_{2j+1}^2} 
\label{eq:q-dep-en}
\end{align}
can be rewritten as the potential~(\ref{eq:interval-massless-appendix}) in the massless case.\footnote{
Let $\alpha$ be real and $b\in\mathbb{Z}$ satisfy $b \leq\alpha\leq b+1$. A useful formula
$\sum_{j\in\mathbb{Z}} \frac{\e^{2\pi i \alpha(j+1/2)}}{A^2+\pi^2(j+1/2)^2}  = (-1)^b \frac{\sinh[(1+2b-2\alpha)A]}{A\cosh(A)}$ can be proved by the residue theorem and contour deformation.
} 

To derive the $\mathcal{O}(m)$ correction to the ground state energy, 
we now compute the chiral condensate in the massless theory.
Upon bosonization it is given as 
\begin{equation}\label{eq:condensate-cosine}
\langle\bar\psi\psi(x)\rangle = - \displaystyle\frac{\e^\gamma g}{2\pi^{3/2}}  \langle 0| :\hspace{-2pt}\cos\big(2\sqrt\pi \phi(x)\big)\hspace{-2pt}:_\infty|0\rangle,
\end{equation}
where the normal ordering~$:\bullet:_\infty$ is taken with respect to the creation and annihilation operators on an infinite line, while $|0\rangle$ is the ground state on the interval.
Let us denote the normal ordering with respect to $a_n$ in~(\ref{interval-an}) by $:\bullet:$.
We find formally
\begin{equation} \label{exp-vev-infinite}
:\hspace{-2pt}\e^{\pm 2\im\sqrt\pi \phi(x)}\hspace{-2pt}:_\infty
=
\e^{\pm 2\im\sqrt\pi \phi(x)}
\exp\left(
\int_0^\infty
 \frac{dk}{\sqrt{\mu^2+ k^2}}
\right) ,
\end{equation}
\begin{align} 
&:\hspace{-2pt}\e^{\pm 2\im\sqrt\pi \phi(x)}\hspace{-2pt}:
=
\e^{\pm 2\im\sqrt\pi \phi(x)}
\nonumber\\
&\qquad\qquad\qquad
\times
\exp\left[
\sum_{n=1}^\infty \frac{2\pi}{L} \frac{\sin^2(k_nx)}{\sqrt{\mu^2+ k_n^2}}
\right].\label{exp-vev-interval}
\end{align}
The right hand sides of~(\ref{exp-vev-infinite}) and~(\ref{exp-vev-interval}) involve UV divergent expressions.
By introducing a cut-off, we obtain the well-defined relation
\begin{align}
:\hspace{-2pt}\e^{\pm 2\im\sqrt\pi \phi(x)}\hspace{-2pt}:_\infty
=
\lambda(x)
:\hspace{-2pt}\e^{\pm 2\im\sqrt\pi \phi(x)}\hspace{-2pt}:, 
\end{align}
where $\lambda(x)$ is defined in (\ref{def:lambda}).
Then~(\ref{eq:condensate-cosine}) gives%
\footnote{%
{With the special values $w_0=1/4$ and $w_1=-1/4$, the chiral condensate~$\langle\bar\psi\psi(x)\rangle$ is finite near the boundaries.} 
}
\begin{align}\label{eq:condensate-general}
&\langle\bar\psi\psi(x)\rangle
=
- \displaystyle\frac{\e^\gamma g}{2\pi^{3/2}} 
\lambda(x)
\nonumber\\
&\qquad\
\times
\cos\left[{2\sqrt{\pi}\phi_0-} 
\sum_{n=1}^\infty \frac{\mu^2\Theta_n}{\omega_n^2}   \sin(k_n x)
\right] .
\end{align}
With this, perturbation theory gives the correction (\ref{eq:energy_m1}) to the ground state energy.

%%%%%%%%%%%%%%%%
\begin{figure*}[th]
\centering\includegraphics[width=12cm]{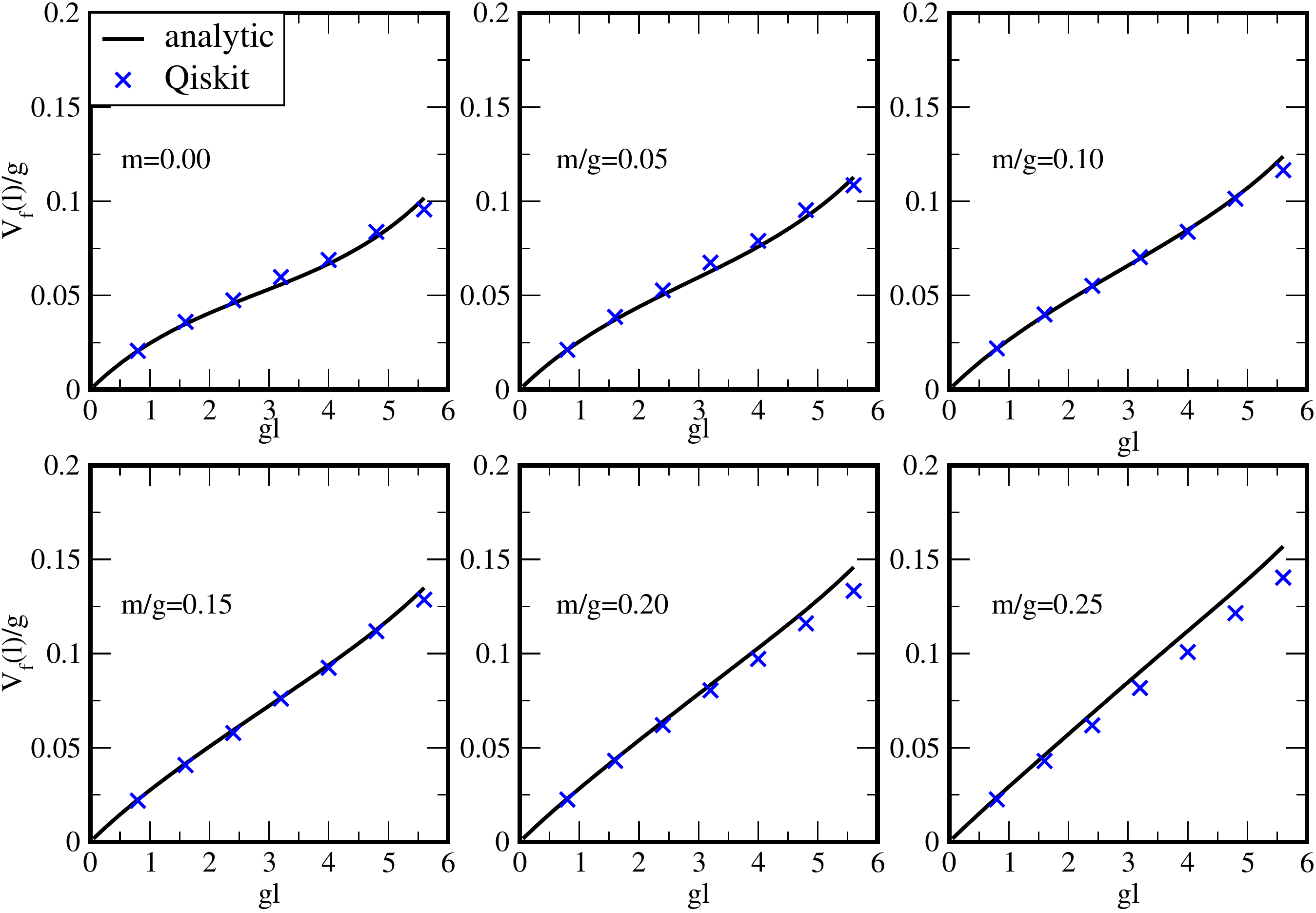}
\caption{Potential~$V_f(\ell)$ for several values of $m/g$, and for $N=15$, $q=0.25$, $\theta_0=0$.
The blue crosses represent the results of quantum simulation without statistical uncertainties.
The black solid curves represent the $\mathcal{O}(m)$ analytic prediction $V_f^{(0)}(\ell)+V_f^{(1)}(\ell)$.
} 
\label{fig:potential-mass}
\end{figure*}
%%%%%%%%%%%%%%%%

For a single probe defined by
\begin{equation}\label{eq:Theta-single-W}
\Theta_\text{probe}(x) :=\left \{
\begin{array}{clc}
\displaystyle \theta_0& \text{ for }x <0 ,\\
\vspace{-4mm} \\
\displaystyle  \theta_0 +2\pi q & \text{ for } x>0,
\end{array}
\right. 
\end{equation}
we find, either from the general formula~(\ref{eq:condensate-general}) or by repeating the derivation, 
\begin{align}\label{eq:condensate-probe}
\langle \bar\psi\psi(x)\rangle_\text{probe}& =  -  \frac{\e^\gamma g}{2\pi^{3/2}}  \cos\Big[\theta_0 +\pi q
\nonumber\\
& \ + \text{sgn}(x)
 \pi q  \left(1-\e^{-\mu|x|}\right)
\Big] \,.
\end{align}
Integrating, over $x\in(-\infty,0]$ for example, the difference between~(\ref{eq:condensate-probe}) and its asymptotic value gives
\begin{align}
& \int_{-\infty}^0 dx    \left(\langle \bar\psi\psi(x)\rangle_\text{probe} +\frac{\e^\gamma g}{2\pi^{3/2}}  \cos\theta_0\right) 
\nonumber\\
&\quad
= \frac{\e^\gamma}{2\pi}
\left[
\cos\theta_0\, {\rm Cin}(\pi q) +\sin\theta_0 {\rm Si}(\pi q)
\right],
\end{align}
where ${\rm Cin}(z)= \int_0^z ds [1-\cos(s)]/s$ and ${\rm Si}(z)= \int_0^z ds \sin(s)/s$.
By considering two well-separated probes, it follows that the constant~$c_q$ in~(\ref{eq:Gross-massive}) is given by~(\ref{eq:cq}).

Finally we consider the large mass limit $m/g\gg 1$ of (\ref{eq:Ham-density-bosonized}).
We assume that $|w_0|, |w_1|<1/2$.
The cosine term forces $\phi$ to be localized near $\phi=0$ in the bulk.
Expanding the cosine in $\phi$, we can write the potential part of $\mathcal{H}_b$ as
\begin{equation}
    C_1
    (\phi+C_2\Theta)^2+\frac{g^2 }{8 \pi ^2}\left(1+\frac{\e^{-\gamma } g}{2 \sqrt{\pi } m}\right)^{-1}\Theta ^2 +\mathcal{O}(\phi^3),
    \nonumber
\end{equation}
where $C_1=\mathcal{O}(m/g)$ and $C_2=\mathcal{O}(g/m)$ are constants.
Substituting~(\ref{eq:Theta-x}) and repeating the analysis leading to~(\ref{eq:ground-state-energy}), we obtain the result~(\ref{eq:large-mass-potential}) for the potential~$V_f(\ell)$.

\subsection{{Volume dependence of the slope of the potential}}
\label{subsec:vol-dep-slope}

\begin{figure}
    \centering
    \includegraphics[scale=0.8]{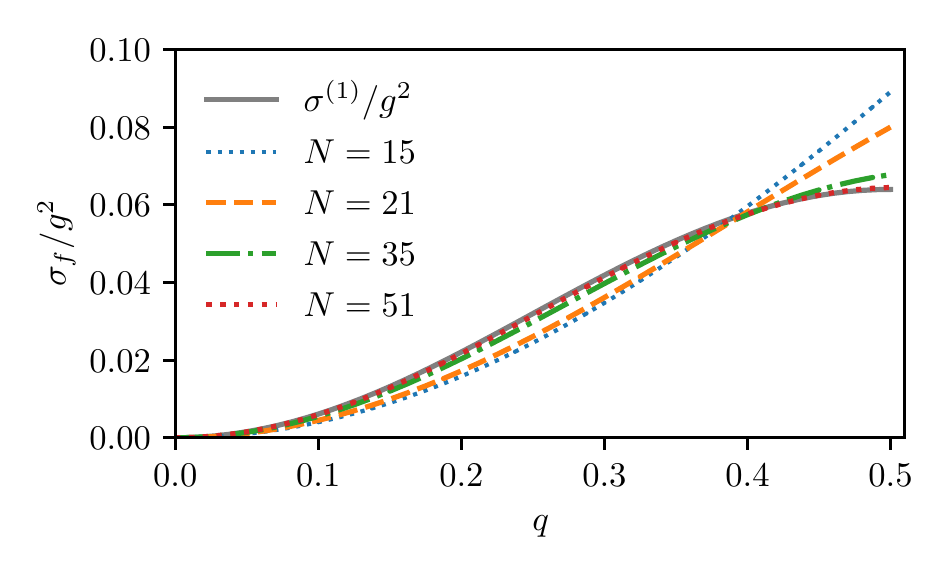}
    \caption{The values of~$(d/d\ell)(V_f^{(0)}+V_f^{(1)})/g^2$ at $\ell=L/2$ for $L=(N-1)a$, $m/g=0.2$, and $ga=0.4$, with the indicated values of $N$. We also plot
    %$\sigma_{\rm Coulomb}/g^2= q^2/2$ and 
    $\sigma^{(1)}/g^2$ given in~(\ref{eq:sigma-1}). }
    \label{fig:sigmaf_midpoint}
\end{figure}

{The results in FIG.~\ref{fig:tension} include finite-size corrections.
We wish to estimate the number of qubits for which such corrections become negligible.
Using our analytic formulas for $V_f^{(0)}+V_f^{(1)}$, we look for the values of volume~$L$ where the slope (the first derivative) {at} an appropriate {point} accurately reproduces the string tension $\sigma^{(1)}$ in the infinite volume.
For~$m/g=0.2$ and $ga=0.4$, we find that with $L={13.6}$  ($N={35}$), the slope at {$\ell=L/2$ reproduces $\sigma^{(1)}$ within 5 percent, as can be seen from Figure~\ref{fig:sigmaf_midpoint}, where $(d/d\ell)(V_f^{(0)}+V_f^{(1)})/g^2$ at $\ell=L/2$ is plotted for several values of $N$.}
We thus expect that an ideal quantum simulation with $N=35$ or larger will exhibit reasonable agreement with the true (infinite-volume) string tension.
}
%%%%%%%%%%%%%%%%%%%%%%%
%%%%%%%%%%%%%%%%%%%%%%%
%%%%%%%%%%%%%%%%%%%%%%%
\section{Mass dependence of the potential}
\label{sec:mass-deps}
%%%%%%%%%%%%%%%%%%%%%%%
%%%%%%%%%%%%%%%%%%%%%%%
%%%%%%%%%%%%%%%%%%%%%%%

Here we study 
the dependence of the potential $V_f(\ell)
$ on the fermion mass~$m$. 
We depict quantum simulation results and $\mathcal{O}(m)$ analytic predictions for $V_f(\ell)$ in FIG.~\ref{fig:potential-mass}~\footnote{{These simulation results include adiabatic errors. See Tables I and II for 
%raw data 
for the systematic errors in similar settings.}}. 
As in Section~\ref{sec:results} 
we take $a=0.4 g^{-1}$, $\delta t =0.3g^{-1}$, $T=99g^{-1}$ and $m_0 =0.5g$.
We use the ``snapshot'' functionality of Qiskit to obtain quantum simulation results without statistical uncertainties.
As we increase the mass within the range $m/g \lesssim 0.15$, the analytic results become less curved. 
At $m/g=0.2$ and $0.25$, the curves 
are almost straight lines.

In the small mass range $0<m/g \lesssim 0.15$, the simulation results match  well the analytic predictions.
For $0.2\lesssim m/g\lesssim 0.25$, the differences between the simulation results and the $\mathcal{O}(m)$ analytic predictions are larger.
If the systematic errors in our results are small,
then this discrepancy should be due to finite $a$ effect or/and breaking of the approximation by the mass perturbation in this regime.

We expect that the latter is a dominant source of the discrepancy as follows.
There is another numerical study \cite{Funcke:2019zna} of the lattice Schwinger model by the density matrix renormalization group (DMRG) method,
which takes the both continuum and infinite-volume limits.
In~\cite{Funcke:2019zna}, analogous deviations of the numerical results from the analytic $\mathcal{O}(m)$ results were found for similar values of $m/g$. 
Since the continuum limit is taken in \cite{Funcke:2019zna},
this implies that 
the mass perturbation theory in the regime $0.2\lesssim m/g\lesssim 0.25$ is no longer reliable at least for large volume.
Although it is nontrivial whether this is still true for the value of our volume,
we regard this as indirect evidence of our expectation. 
Then our results in FIG.~\ref{fig:potential-mass} likely give a prediction for the middle range of mass between small mass and massive limit.  
It would be important to check the consistency of the results between the quantum computing and the DMRG in such a non-perturbative mass regime in the future.

%%%%%%%%%%%%%%%%%%%%%%%
%%%%%%%%%%%%%%%%%%%%%%%
%%%%%%%%%%%%%%%%%%%%%%%
\bibliographystyle{utphys}
\bibliography{quantum_computation}

\providecommand{\href}[2]{#2}\begingroup\raggedright\begin{thebibliography}{10}

\bibitem{Wilson:1974sk}
K.~G. Wilson, ``Confinement of quarks,''
  \href{http://dx.doi.org/10.1103/PhysRevD.10.2445}{{\em Phys. Rev. D}
  {\bfseries 10} (1974) 2445--2459}.

\bibitem{Creutz:1980zw}
M.~Creutz, ``{Monte Carlo Study of Quantized SU(2) Gauge Theory},''
  \href{http://dx.doi.org/10.1103/PhysRevD.21.2308}{{\em Phys. Rev. D}
  {\bfseries 21} (1980) 2308--2315}.

\bibitem{Otto:1984qr}
S.~W. Otto and J.~D. Stack, ``{The SU(3) Heavy Quark Potential with High
  Statistics},'' \href{http://dx.doi.org/10.1103/PhysRevLett.52.2328}{{\em
  Phys. Rev. Lett.} {\bfseries 52} (1984) 2328}.

\bibitem{Bali:2000gf}
G.~S. Bali, ``{QCD forces and heavy quark bound states},''
  \href{http://dx.doi.org/10.1016/S0370-1573(00)00079-X}{{\em Phys. Rept.}
  {\bfseries 343} (2001) 1--136},
  \href{http://arxiv.org/abs/hep-ph/0001312}{{\ttfamily arXiv:hep-ph/0001312}}.

\bibitem{Eichten:1974af}
E.~Eichten, K.~Gottfried, T.~Kinoshita, J.~B. Kogut, K.~D. Lane, and T.-M. Yan,
  ``{The Spectrum of Charmonium},''
  \href{http://dx.doi.org/10.1103/PhysRevLett.34.369}{{\em Phys. Rev. Lett.}
  {\bfseries 34} (1975) 369--372}. [Erratum: Phys.Rev.Lett. 36, 1276 (1976)].

\bibitem{Aarts:2015tyj}
G.~Aarts, ``{Introductory lectures on lattice QCD at nonzero baryon number},''
  \href{http://dx.doi.org/10.1088/1742-6596/706/2/022004}{{\em J. Phys. Conf.
  Ser.} {\bfseries 706} no.~2, (2016) 022004},
\href{http://arxiv.org/abs/1512.05145}{{\ttfamily arXiv:1512.05145 [hep-lat]}}.
%%CITATION = ARXIV:1512.05145;%%.

\bibitem{Schwinger:1962tp}
J.~S. Schwinger, ``{Gauge Invariance and Mass. 2.},''
  \href{http://dx.doi.org/10.1103/PhysRev.128.2425}{{\em Phys. Rev.} {\bfseries
  128} (1962) 2425--2429}.

\bibitem{Coleman:1975pw}
S.~R. Coleman, R.~Jackiw, and L.~Susskind, ``{Charge Shielding and Quark
  Confinement in the Massive Schwinger Model},''
\href{http://dx.doi.org/10.1016/0003-4916(75)90212-2}{{\em Annals Phys.}
  {\bfseries 93} (1975) 267}.
%%CITATION = APNYA,93,267;%%.

\bibitem{Gross:1995bp}
D.~J. Gross, I.~R. Klebanov, A.~V. Matytsin, and A.~V. Smilga, ``{Screening
  versus confinement in (1+1)-dimensions},''
  \href{http://dx.doi.org/10.1016/0550-3213(95)00655-9}{{\em Nucl. Phys.}
  {\bfseries B461} (1996) 109--130},
\href{http://arxiv.org/abs/hep-th/9511104}{{\ttfamily arXiv:hep-th/9511104
  [hep-th]}}.
%%CITATION = HEP-TH/9511104;%%.

\bibitem{Preskill2018quantumcomputingin}
J.~Preskill, ``Quantum {C}omputing in the {NISQ} era and beyond,''
  \href{http://dx.doi.org/10.22331/q-2018-08-06-79}{{\em {Quantum}} {\bfseries
  2} (Aug., 2018) 79}. \url{https://doi.org/10.22331/q-2018-08-06-79}.

\bibitem{Martinez:2016yna}
E.~A. Martinez {\em et~al.}, ``{Real-time dynamics of lattice gauge theories
  with a few-qubit quantum computer},''
  \href{http://dx.doi.org/10.1038/nature18318}{{\em Nature} {\bfseries 534}
  (2016) 516--519},
\href{http://arxiv.org/abs/1605.04570}{{\ttfamily arXiv:1605.04570
  [quant-ph]}}.
%%CITATION = ARXIV:1605.04570;%%.

\bibitem{Muschik:2016tws}
C.~Muschik, M.~Heyl, E.~Martinez, T.~Monz, P.~Schindler, B.~Vogell,
  M.~Dalmonte, P.~Hauke, R.~Blatt, and P.~Zoller, ``{U(1) Wilson lattice gauge
  theories in digital quantum simulators},''
  \href{http://dx.doi.org/10.1088/1367-2630/aa89ab}{{\em New J. Phys.}
  {\bfseries 19} no.~10, (2017) 103020},
\href{http://arxiv.org/abs/1612.08653}{{\ttfamily arXiv:1612.08653
  [quant-ph]}}.
%%CITATION = ARXIV:1612.08653;%%.

\bibitem{Klco:2018kyo}
N.~Klco, E.~F. Dumitrescu, A.~J. McCaskey, T.~D. Morris, R.~C. Pooser, M.~Sanz,
  E.~Solano, P.~Lougovski, and M.~J. Savage, ``{Quantum-classical computation
  of Schwinger model dynamics using quantum computers},''
  \href{http://dx.doi.org/10.1103/PhysRevA.98.032331}{{\em Phys. Rev.}
  {\bfseries A98} no.~3, (2018) 032331},
\href{http://arxiv.org/abs/1803.03326}{{\ttfamily arXiv:1803.03326
  [quant-ph]}}.
%%CITATION = ARXIV:1803.03326;%%.

\bibitem{Kokail:2018eiw}
C.~Kokail {\em et~al.}, ``{Self-verifying variational quantum simulation of
  lattice models},'' \href{http://dx.doi.org/10.1038/s41586-019-1177-4}{{\em
  Nature} {\bfseries 569} no.~7756, (2019) 355--360},
\href{http://arxiv.org/abs/1810.03421}{{\ttfamily arXiv:1810.03421
  [quant-ph]}}.
%%CITATION = ARXIV:1810.03421;%%.

\bibitem{Magnifico:2019kyj}
G.~Magnifico, M.~Dalmonte, P.~Facchi, S.~Pascazio, F.~V. Pepe, and
  E.~Ercolessi, ``{Real Time Dynamics and Confinement in the $\mathbb{Z}_{n}$
  Schwinger-Weyl lattice model for 1+1 QED},''
\href{http://arxiv.org/abs/1909.04821}{{\ttfamily arXiv:1909.04821
  [quant-ph]}}.
%%CITATION = ARXIV:1909.04821;%%.

\bibitem{Chakraborty:2020uhf}
B.~Chakraborty, M.~Honda, T.~Izubuchi, Y.~Kikuchi, and A.~Tomiya, ``{Digital
  Quantum Simulation of the Schwinger Model with Topological Term via Adiabatic
  State Preparation},'' \href{http://arxiv.org/abs/2001.00485}{{\ttfamily
  arXiv:2001.00485 [hep-lat]}}.

\bibitem{Yamamoto:2021vxp}
A.~Yamamoto, ``{Quantum variational approach to lattice gauge theory at nonzero
  density},'' \href{http://arxiv.org/abs/2104.10669}{{\ttfamily
  arXiv:2104.10669 [hep-lat]}}.

\bibitem{2020PhRvR...2b3342K}
D.~E. {Kharzeev} and Y.~{Kikuchi}, ``{Real-time chiral dynamics from a digital
  quantum simulation},''
  \href{http://dx.doi.org/10.1103/PhysRevResearch.2.023342}{{\em Physical
  Review Research} {\bfseries 2} no.~2, (June, 2020) 023342},
  \href{http://arxiv.org/abs/2001.00698}{{\ttfamily arXiv:2001.00698
  [hep-ph]}}.

\bibitem{2021arXiv210608394D}
W.~A. {de Jong}, K.~{Lee}, J.~{Mulligan}, M.~{P{\l}osko{\'n}}, F.~{Ringer}, and
  X.~{Yao}, ``{Quantum simulation of non-equilibrium dynamics and
  thermalization in the Schwinger model},'' {\em arXiv e-prints} (June, 2021)
  arXiv:2106.08394, \href{http://arxiv.org/abs/2106.08394}{{\ttfamily
  arXiv:2106.08394 [quant-ph]}}.

\bibitem{Bernien_2017}
H.~Bernien {\em et~al.}, ``Probing many-body dynamics on a 51-atom quantum
  simulator,'' \href{http://dx.doi.org/10.1038/nature24622}{{\em Nature}
  {\bfseries 551} no.~7682, (Nov, 2017) 579–584},
  \href{http://arxiv.org/abs/1707.04344}{{\ttfamily arXiv:1707.04344
  [quant-ph]}}. \url{http://dx.doi.org/10.1038/nature24622}.

\bibitem{Surace_2020}
F.~M. Surace, P.~P. Mazza, G.~Giudici, A.~Lerose, A.~Gambassi, and M.~Dalmonte,
  ``Lattice gauge theories and string dynamics in rydberg atom quantum
  simulators,'' \href{http://dx.doi.org/10.1103/physrevx.10.021041}{{\em
  Physical Review X} {\bfseries 10} no.~2, (May, 2020) 021041}.

\bibitem{Jordan:2011ne}
S.~P. Jordan, K.~S.~M. Lee, and J.~Preskill, ``{Quantum Algorithms for Quantum
  Field Theories},'' \href{http://dx.doi.org/10.1126/science.1217069}{{\em
  Science} {\bfseries 336} (2012) 1130--1133},
\href{http://arxiv.org/abs/1111.3633}{{\ttfamily arXiv:1111.3633 [quant-ph]}}.
%%CITATION = ARXIV:1111.3633;%%.

\bibitem{Jordan:2011ci}
S.~P. Jordan, K.~S.~M. Lee, and J.~Preskill, ``{Quantum Computation of
  Scattering in Scalar Quantum Field Theories},''
  \href{http://arxiv.org/abs/1112.4833}{{\ttfamily arXiv:1112.4833 [hep-th]}}.
[Quant. Inf. Comput.14,1014(2014)].
%%CITATION = ARXIV:1112.4833;%%.

\bibitem{Jordan:2014tma}
S.~P. Jordan, K.~S.~M. Lee, and J.~Preskill, ``{Quantum Algorithms for
  Fermionic Quantum Field Theories},''
\href{http://arxiv.org/abs/1404.7115}{{\ttfamily arXiv:1404.7115 [hep-th]}}.
%%CITATION = ARXIV:1404.7115;%%.

\bibitem{Garcia-Alvarez:2014uda}
L.~Garcia-Alvarez, J.~Casanova, A.~Mezzacapo, I.~L. Egusquiza, L.~Lamata,
  G.~Romero, and E.~Solano, ``{Fermion-Fermion Scattering in Quantum Field
  Theory with Superconducting Circuits},''
  \href{http://dx.doi.org/10.1103/PhysRevLett.114.070502}{{\em Phys. Rev.
  Lett.} {\bfseries 114} no.~7, (2015) 070502},
\href{http://arxiv.org/abs/1404.2868}{{\ttfamily arXiv:1404.2868 [quant-ph]}}.
%%CITATION = ARXIV:1404.2868;%%.

\bibitem{Wiese:2014rla}
U.-J. Wiese, ``{Towards Quantum Simulating QCD},''
  \href{http://dx.doi.org/10.1016/j.nuclphysa.2014.09.102}{{\em Nucl. Phys.}
  {\bfseries A931} (2014) 246--256},
\href{http://arxiv.org/abs/1409.7414}{{\ttfamily arXiv:1409.7414 [hep-th]}}.
%%CITATION = ARXIV:1409.7414;%%.

\bibitem{Marcos:2014lda}
D.~Marcos, P.~Widmer, E.~Rico, M.~Hafezi, P.~Rabl, U.~J. Wiese, and P.~Zoller,
  ``{Two-dimensional Lattice Gauge Theories with Superconducting Quantum
  Circuits},'' \href{http://dx.doi.org/10.1016/j.aop.2014.09.011}{{\em Annals
  Phys.} {\bfseries 351} (2014) 634--654},
\href{http://arxiv.org/abs/1407.6066}{{\ttfamily arXiv:1407.6066 [quant-ph]}}.
%%CITATION = ARXIV:1407.6066;%%.

\bibitem{Mezzacapo:2015bra}
A.~Mezzacapo, E.~Rico, C.~Sabin, I.~L. Egusquiza, L.~Lamata, and E.~Solano,
  ``{Non-Abelian $SU(2)$ Lattice Gauge Theories in Superconducting Circuits},''
  \href{http://dx.doi.org/10.1103/PhysRevLett.115.240502}{{\em Phys. Rev.
  Lett.} {\bfseries 115} no.~24, (2015) 240502},
\href{http://arxiv.org/abs/1505.04720}{{\ttfamily arXiv:1505.04720
  [quant-ph]}}.
%%CITATION = ARXIV:1505.04720;%%.

\bibitem{Macridin:2018gdw}
A.~Macridin, P.~Spentzouris, J.~Amundson, and R.~Harnik, ``{Electron-Phonon
  Systems on a Universal Quantum Computer},''
  \href{http://dx.doi.org/10.1103/PhysRevLett.121.110504}{{\em Phys. Rev.
  Lett.} {\bfseries 121} no.~11, (2018) 110504},
\href{http://arxiv.org/abs/1802.07347}{{\ttfamily arXiv:1802.07347
  [quant-ph]}}.
%%CITATION = ARXIV:1802.07347;%%.

\bibitem{Lamm:2018siq}
H.~Lamm and S.~Lawrence, ``{Simulation of Nonequilibrium Dynamics on a Quantum
  Computer},'' \href{http://dx.doi.org/10.1103/PhysRevLett.121.170501}{{\em
  Phys. Rev. Lett.} {\bfseries 121} no.~17, (2018) 170501},
\href{http://arxiv.org/abs/1806.06649}{{\ttfamily arXiv:1806.06649
  [quant-ph]}}.
%%CITATION = ARXIV:1806.06649;%%.

\bibitem{Klco:2018zqz}
N.~Klco and M.~J. Savage, ``{Digitization of scalar fields for quantum
  computing},'' \href{http://dx.doi.org/10.1103/PhysRevA.99.052335}{{\em Phys.
  Rev.} {\bfseries A99} no.~5, (2019) 052335},
\href{http://arxiv.org/abs/1808.10378}{{\ttfamily arXiv:1808.10378
  [quant-ph]}}.
%%CITATION = ARXIV:1808.10378;%%.

\bibitem{Gustafson:2019mpk}
E.~Gustafson, Y.~Meurice, and J.~Unmuth-Yockey, ``{Quantum simulation of
  scattering in the quantum Ising model},''
  \href{http://dx.doi.org/10.1103/PhysRevD.99.094503}{{\em Phys. Rev.}
  {\bfseries D99} no.~9, (2019) 094503},
\href{http://arxiv.org/abs/1901.05944}{{\ttfamily arXiv:1901.05944 [hep-lat]}}.
%%CITATION = ARXIV:1901.05944;%%.

\bibitem{Alexandru:2019ozf}
{\bfseries NuQS} Collaboration, A.~Alexandru, P.~F. Bedaque, H.~Lamm, and
  S.~Lawrence, ``{$\sigma$ Models on Quantum Computers},''
  \href{http://dx.doi.org/10.1103/PhysRevLett.123.090501}{{\em Phys. Rev.
  Lett.} {\bfseries 123} no.~9, (2019) 090501},
\href{http://arxiv.org/abs/1903.06577}{{\ttfamily arXiv:1903.06577 [hep-lat]}}.
%%CITATION = ARXIV:1903.06577;%%.

\bibitem{Klco:2019xro}
N.~Klco and M.~J. Savage, ``{Minimally-Entangled State Preparation of Localized
  Wavefunctions on Quantum Computers},''
\href{http://arxiv.org/abs/1904.10440}{{\ttfamily arXiv:1904.10440
  [quant-ph]}}.
%%CITATION = ARXIV:1904.10440;%%.

\bibitem{Klco:2019evd}
N.~Klco, J.~R. Stryker, and M.~J. Savage, ``{SU(2) non-Abelian gauge field
  theory in one dimension on digital quantum computers},''
\href{http://arxiv.org/abs/1908.06935}{{\ttfamily arXiv:1908.06935
  [quant-ph]}}.
%%CITATION = ARXIV:1908.06935;%%.

\bibitem{Lamm:2019uyc}
{\bfseries NuQS} Collaboration, H.~Lamm, S.~Lawrence, and Y.~Yamauchi,
  ``{Parton Physics on a Quantum Computer},''
\href{http://arxiv.org/abs/1908.10439}{{\ttfamily arXiv:1908.10439 [hep-lat]}}.
%%CITATION = ARXIV:1908.10439;%%.

\bibitem{Mueller:2019qqj}
N.~Mueller, A.~Tarasov, and R.~Venugopalan, ``{Deeply inelastic scattering
  structure functions on a hybrid quantum computer},''
\href{http://arxiv.org/abs/1908.07051}{{\ttfamily arXiv:1908.07051 [hep-th]}}.
%%CITATION = ARXIV:1908.07051;%%.

\bibitem{Gustafson:2019vsd}
E.~Gustafson, P.~Dreher, Z.~Hang, and Y.~Meurice, ``{Real time evolution of a
  one-dimensional field theory on a 20 qubit machine},''
\href{http://arxiv.org/abs/1910.09478}{{\ttfamily arXiv:1910.09478 [hep-lat]}}.
%%CITATION = ARXIV:1910.09478;%%.

\bibitem{2020arXiv200615746B}
A.~J. {Buser}, T.~{Bhattacharya}, L.~{Cincio}, and R.~{Gupta}, ``{State
  preparation and measurement in a quantum simulation of the O(3) sigma
  model},'' {\em arXiv e-prints} (June, 2020) arXiv:2006.15746,
  \href{http://arxiv.org/abs/2006.15746}{{\ttfamily arXiv:2006.15746
  [quant-ph]}}.

\bibitem{2020PhRvL.125p0503M}
F.~{Mei}, Q.~{Guo}, Y.-F. {Yu}, L.~{Xiao}, S.-L. {Zhu}, and S.~{Jia},
  ``{Digital Simulation of Topological Matter on Programmable Quantum
  Processors},'' \href{http://dx.doi.org/10.1103/PhysRevLett.125.160503}{{\em
  \prl} {\bfseries 125} no.~16, (Oct., 2020) 160503},
  \href{http://arxiv.org/abs/2003.06086}{{\ttfamily arXiv:2003.06086
  [quant-ph]}}.

\bibitem{2020arXiv201007965A}
F.~{Arute} {\em et~al.}, ``{Observation of separated dynamics of charge and
  spin in the Fermi-Hubbard model},'' {\em arXiv e-prints} (Oct., 2020)
  arXiv:2010.07965, \href{http://arxiv.org/abs/2010.07965}{{\ttfamily
  arXiv:2010.07965 [quant-ph]}}.

\bibitem{2020arXiv201106576B}
A.~J. {Buser}, H.~{Gharibyan}, M.~{Hanada}, M.~{Honda}, and J.~{Liu},
  ``{Quantum simulation of gauge theory via orbifold lattice},'' {\em arXiv
  e-prints} (Nov., 2020) arXiv:2011.06576,
  \href{http://arxiv.org/abs/2011.06576}{{\ttfamily arXiv:2011.06576
  [hep-th]}}.

\bibitem{2020PhRvA.102e2422K}
N.~{Klco} and M.~J. {Savage}, ``{Fixed-point quantum circuits for quantum field
  theories},'' \href{http://dx.doi.org/10.1103/PhysRevA.102.052422}{{\em \pra}
  {\bfseries 102} no.~5, (Nov., 2020) 052422},
  \href{http://arxiv.org/abs/2002.02018}{{\ttfamily arXiv:2002.02018
  [quant-ph]}}.

\bibitem{2020JHEP...12..011L}
J.~{Liu} and Y.~{Xin}, ``{Quantum simulation of quantum field theories as
  quantum chemistry},'' \href{http://dx.doi.org/10.1007/JHEP12(2020)011}{{\em
  Journal of High Energy Physics} {\bfseries 2020} no.~12, (Dec., 2020) 11},
  \href{http://arxiv.org/abs/2004.13234}{{\ttfamily arXiv:2004.13234
  [hep-th]}}.

\bibitem{2020arXiv201209194S}
Y.~{Su}, H.-Y. {Huang}, and E.~T. {Campbell}, ``{Nearly tight Trotterization of
  interacting electrons},'' {\em arXiv e-prints} (Dec., 2020) arXiv:2012.09194,
  \href{http://arxiv.org/abs/2012.09194}{{\ttfamily arXiv:2012.09194
  [quant-ph]}}.

\bibitem{2021PhRvD.103e4507G}
E.~J. {Gustafson} and H.~{Lamm}, ``{Toward quantum simulations of Z$_{2}$ gauge
  theory without state preparation},''
  \href{http://dx.doi.org/10.1103/PhysRevD.103.054507}{{\em \prd} {\bfseries
  103} no.~5, (Mar., 2021) 054507},
  \href{http://arxiv.org/abs/2011.11677}{{\ttfamily arXiv:2011.11677
  [hep-lat]}}.

\bibitem{2021PhRvA.103d2410B}
J.~{Barata}, N.~{Mueller}, A.~{Tarasov}, and R.~{Venugopalan},
  ``{Single-particle digitization strategy for quantum computation of a
  {\ensuremath{\phi}}$^{4}$ scalar field theory},''
  \href{http://dx.doi.org/10.1103/PhysRevA.103.042410}{{\em \pra} {\bfseries
  103} no.~4, (Apr., 2021) 042410},
  \href{http://arxiv.org/abs/2012.00020}{{\ttfamily arXiv:2012.00020
  [hep-th]}}.

\bibitem{2021PhRvD.103i4501C}
A.~{Ciavarella}, N.~{Klco}, and M.~J. {Savage}, ``{Trailhead for quantum
  simulation of SU(3) Yang-Mills lattice gauge theory in the local multiplet
  basis},'' \href{http://dx.doi.org/10.1103/PhysRevD.103.094501}{{\em \prd}
  {\bfseries 103} no.~9, (May, 2021) 094501},
  \href{http://arxiv.org/abs/2101.10227}{{\ttfamily arXiv:2101.10227
  [quant-ph]}}.

\bibitem{2021JHEP...07..140G}
H.~{Gharibyan}, M.~{Hanada}, M.~{Honda}, and J.~{Liu}, ``{Toward simulating
  superstring/M-theory on a quantum computer},''
  \href{http://dx.doi.org/10.1007/JHEP07(2021)140}{{\em Journal of High Energy
  Physics} {\bfseries 2021} no.~7, (July, 2021) 140},
  \href{http://arxiv.org/abs/2011.06573}{{\ttfamily arXiv:2011.06573
  [hep-th]}}.

\bibitem{2021PhRvD.104a4512E}
M.~G. {Echevarria}, I.~L. {Egusquiza}, E.~{Rico}, and G.~{Schnell}, ``{Quantum
  simulation of light-front parton correlators},''
  \href{http://dx.doi.org/10.1103/PhysRevD.104.014512}{{\em \prd} {\bfseries
  104} no.~1, (July, 2021) 014512},
  \href{http://arxiv.org/abs/2011.01275}{{\ttfamily arXiv:2011.01275
  [quant-ph]}}.

\bibitem{2021PRXQ....2c0334P}
D.~{Paulson}, L.~{Dellantonio}, J.~F. {Haase}, A.~{Celi}, A.~{Kan}, A.~{Jena},
  C.~{Kokail}, R.~{van Bijnen}, K.~{Jansen}, P.~{Zoller}, and C.~A. {Muschik},
  ``{Simulating 2D Effects in Lattice Gauge Theories on a Quantum Computer},''
  \href{http://dx.doi.org/10.1103/PRXQuantum.2.030334}{{\em PRX Quantum}
  {\bfseries 2} no.~3, (Aug., 2021) 030334},
  \href{http://arxiv.org/abs/2008.09252}{{\ttfamily arXiv:2008.09252
  [quant-ph]}}.

\bibitem{2021PhRvD.104h6013L}
J.~{Liu} and Y.-Z. {Li}, ``{Quantum simulation of cosmic inflation},''
  \href{http://dx.doi.org/10.1103/PhysRevD.104.086013}{{\em \prd} {\bfseries
  104} no.~8, (Oct., 2021) 086013},
  \href{http://arxiv.org/abs/2009.10921}{{\ttfamily arXiv:2009.10921
  [quant-ph]}}.

\bibitem{2021PhRvD.104g4505D}
Z.~{Davoudi}, I.~{Raychowdhury}, and A.~{Shaw}, ``{Search for efficient
  formulations for Hamiltonian simulation of non-Abelian lattice gauge
  theories},'' \href{http://dx.doi.org/10.1103/PhysRevD.104.074505}{{\em \prd}
  {\bfseries 104} no.~7, (Oct., 2021) 074505},
  \href{http://arxiv.org/abs/2009.11802}{{\ttfamily arXiv:2009.11802
  [hep-lat]}}.

\bibitem{2021PhRvL.127u2001B}
C.~W. {Bauer}, B.~{Nachman}, and M.~{Freytsis}, ``{Simulating Collider Physics
  on Quantum Computers Using Effective Field Theories},''
  \href{http://dx.doi.org/10.1103/PhysRevLett.127.212001}{{\em \prl} {\bfseries
  127} no.~21, (Nov., 2021) 212001},
  \href{http://arxiv.org/abs/2102.05044}{{\ttfamily arXiv:2102.05044
  [hep-ph]}}.

\bibitem{Zohar:2012ay}
E.~Zohar, J.~I. Cirac, and B.~Reznik, ``{Simulating Compact Quantum
  Electrodynamics with ultracold atoms: Probing confinement and nonperturbative
  effects},'' \href{http://dx.doi.org/10.1103/PhysRevLett.109.125302}{{\em
  Phys. Rev. Lett.} {\bfseries 109} (2012) 125302},
\href{http://arxiv.org/abs/1204.6574}{{\ttfamily arXiv:1204.6574 [quant-ph]}}.
%%CITATION = ARXIV:1204.6574;%%.

\bibitem{Banerjee:2012pg}
D.~Banerjee, M.~Dalmonte, M.~Muller, E.~Rico, P.~Stebler, U.~J. Wiese, and
  P.~Zoller, ``{Atomic Quantum Simulation of Dynamical Gauge Fields coupled to
  Fermionic Matter: From String Breaking to Evolution after a Quench},''
  \href{http://dx.doi.org/10.1103/PhysRevLett.109.175302}{{\em Phys. Rev.
  Lett.} {\bfseries 109} (2012) 175302},
\href{http://arxiv.org/abs/1205.6366}{{\ttfamily arXiv:1205.6366
  [cond-mat.quant-gas]}}.
%%CITATION = ARXIV:1205.6366;%%.

\bibitem{Zohar:2012xf}
E.~Zohar, J.~I. Cirac, and B.~Reznik, ``{Cold-Atom Quantum Simulator for SU(2)
  Yang-Mills Lattice Gauge Theory},''
  \href{http://dx.doi.org/10.1103/PhysRevLett.110.125304}{{\em Phys. Rev.
  Lett.} {\bfseries 110} no.~12, (2013) 125304},
\href{http://arxiv.org/abs/1211.2241}{{\ttfamily arXiv:1211.2241 [quant-ph]}}.
%%CITATION = ARXIV:1211.2241;%%.

\bibitem{Banerjee:2012xg}
D.~Banerjee, M.~Bogli, M.~Dalmonte, E.~Rico, P.~Stebler, U.~J. Wiese, and
  P.~Zoller, ``{Atomic Quantum Simulation of U(N) and SU(N) Non-Abelian Lattice
  Gauge Theories},''
  \href{http://dx.doi.org/10.1103/PhysRevLett.110.125303}{{\em Phys. Rev.
  Lett.} {\bfseries 110} no.~12, (2013) 125303},
\href{http://arxiv.org/abs/1211.2242}{{\ttfamily arXiv:1211.2242
  [cond-mat.quant-gas]}}.
%%CITATION = ARXIV:1211.2242;%%.

\bibitem{Wiese:2013uua}
U.-J. Wiese, ``{Ultracold Quantum Gases and Lattice Systems: Quantum Simulation
  of Lattice Gauge Theories},''
  \href{http://dx.doi.org/10.1002/andp.201300104}{{\em Annalen Phys.}
  {\bfseries 525} (2013) 777--796},
\href{http://arxiv.org/abs/1305.1602}{{\ttfamily arXiv:1305.1602 [quant-ph]}}.
%%CITATION = ARXIV:1305.1602;%%.

\bibitem{Zohar:2015hwa}
E.~Zohar, J.~I. Cirac, and B.~Reznik, ``{Quantum Simulations of Lattice Gauge
  Theories using Ultracold Atoms in Optical Lattices},''
  \href{http://dx.doi.org/10.1088/0034-4885/79/1/014401}{{\em Rept. Prog.
  Phys.} {\bfseries 79} no.~1, (2016) 014401},
\href{http://arxiv.org/abs/1503.02312}{{\ttfamily arXiv:1503.02312
  [quant-ph]}}.
%%CITATION = ARXIV:1503.02312;%%.

\bibitem{Bazavov:2015kka}
A.~Bazavov, Y.~Meurice, S.-W. Tsai, J.~Unmuth-Yockey, and J.~Zhang,
  ``{Gauge-invariant implementation of the Abelian Higgs model on optical
  lattices},'' \href{http://dx.doi.org/10.1103/PhysRevD.92.076003}{{\em Phys.
  Rev.} {\bfseries D92} no.~7, (2015) 076003},
\href{http://arxiv.org/abs/1503.08354}{{\ttfamily arXiv:1503.08354 [hep-lat]}}.
%%CITATION = ARXIV:1503.08354;%%.

\bibitem{Zohar:2016iic}
E.~Zohar, A.~Farace, B.~Reznik, and J.~I. Cirac, ``{Digital lattice gauge
  theories},'' \href{http://dx.doi.org/10.1103/PhysRevA.95.023604}{{\em Phys.
  Rev.} {\bfseries A95} no.~2, (2017) 023604},
\href{http://arxiv.org/abs/1607.08121}{{\ttfamily arXiv:1607.08121
  [quant-ph]}}.
%%CITATION = ARXIV:1607.08121;%%.

\bibitem{Bermudez:2017yrq}
A.~Bermudez, G.~Aarts, and M.~Muller, ``{Quantum sensors for the generating
  functional of interacting quantum field theories},''
  \href{http://dx.doi.org/10.1103/PhysRevX.7.041012}{{\em Phys. Rev.}
  {\bfseries X7} no.~4, (2017) 041012},
\href{http://arxiv.org/abs/1704.02877}{{\ttfamily arXiv:1704.02877
  [quant-ph]}}.
%%CITATION = ARXIV:1704.02877;%%.

\bibitem{Zache:2018jbt}
T.~V. Zache, F.~Hebenstreit, F.~Jendrzejewski, M.~K. Oberthaler, J.~Berges, and
  P.~Hauke, ``{Quantum simulation of lattice gauge theories using Wilson
  fermions},'' \href{http://dx.doi.org/10.1088/2058-9565/aac33b}{{\em Sci.
  Technol.} {\bfseries 3} (2018) 034010},
\href{http://arxiv.org/abs/1802.06704}{{\ttfamily arXiv:1802.06704
  [cond-mat.quant-gas]}}.
%%CITATION = ARXIV:1802.06704;%%.

\bibitem{Zhang:2018ufj}
J.~Zhang, J.~Unmuth-Yockey, J.~Zeiher, A.~Bazavov, S.~W. Tsai, and Y.~Meurice,
  ``{Quantum simulation of the universal features of the Polyakov loop},''
  \href{http://dx.doi.org/10.1103/PhysRevLett.121.223201}{{\em Phys. Rev.
  Lett.} {\bfseries 121} no.~22, (2018) 223201},
\href{http://arxiv.org/abs/1803.11166}{{\ttfamily arXiv:1803.11166 [hep-lat]}}.
%%CITATION = ARXIV:1803.11166;%%.

\bibitem{Lu:2018pjk}
H.-H. Lu {\em et~al.}, ``{Simulations of Subatomic Many-Body Physics on a
  Quantum Frequency Processor},''
  \href{http://dx.doi.org/10.1103/PhysRevA.100.012320}{{\em Phys. Rev.}
  {\bfseries A100} no.~1, (2019) 012320},
\href{http://arxiv.org/abs/1810.03959}{{\ttfamily arXiv:1810.03959
  [quant-ph]}}.
%%CITATION = ARXIV:1810.03959;%%.

\bibitem{Roy:2020ppa}
A.~Roy, D.~Schuricht, J.~Hauschild, F.~Pollmann, and H.~Saleur, ``{The quantum
  sine-Gordon model with quantum circuits},''
  \href{http://arxiv.org/abs/2007.06874}{{\ttfamily arXiv:2007.06874
  [quant-ph]}}.

\bibitem{2020arXiv200616248T}
M.~C. {Tran}, Y.~{Su}, D.~{Carney}, and J.~M. {Taylor}, ``{Faster Digital
  Quantum Simulation by Symmetry Protection},'' {\em arXiv e-prints} (June,
  2020) arXiv:2006.16248, \href{http://arxiv.org/abs/2006.16248}{{\ttfamily
  arXiv:2006.16248 [quant-ph]}}.

\bibitem{2020PhRvD.101k4502R}
I.~{Raychowdhury} and J.~R. {Stryker}, ``{Loop, string, and hadron dynamics in
  SU(2) Hamiltonian lattice gauge theories},''
  \href{http://dx.doi.org/10.1103/PhysRevD.101.114502}{{\em \prd} {\bfseries
  101} no.~11, (June, 2020) 114502},
  \href{http://arxiv.org/abs/1912.06133}{{\ttfamily arXiv:1912.06133
  [hep-lat]}}.

\bibitem{2021PhRvD.103k4505G}
E.~J. {Gustafson}, ``{Prospects for simulating a qudit-based model of (1 +1 )D
  scalar QED},'' \href{http://dx.doi.org/10.1103/PhysRevD.103.114505}{{\em
  \prd} {\bfseries 103} no.~11, (June, 2021) 114505},
  \href{http://arxiv.org/abs/2104.10136}{{\ttfamily arXiv:2104.10136
  [quant-ph]}}.

\bibitem{Banks:1975gq}
T.~Banks, L.~Susskind, and J.~B. Kogut, ``{Strong Coupling Calculations of
  Lattice Gauge Theories: (1+1)-Dimensional Exercises},''
  \href{http://dx.doi.org/10.1103/PhysRevD.13.1043}{{\em Phys. Rev. D}
  {\bfseries 13} (1976) 1043}.

\bibitem{Hamer:1997dx}
C.~J. Hamer, W.-h. Zheng, and J.~Oitmaa, ``{Series expansions for the massive
  Schwinger model in Hamiltonian lattice theory},''
  \href{http://dx.doi.org/10.1103/PhysRevD.56.55}{{\em Phys. Rev.} {\bfseries
  D56} (1997) 55--67},
\href{http://arxiv.org/abs/hep-lat/9701015}{{\ttfamily arXiv:hep-lat/9701015
  [hep-lat]}}.
%%CITATION = HEP-LAT/9701015;%%.

\bibitem{Kogut:1974ag}
J.~B. Kogut and L.~Susskind, ``{Hamiltonian Formulation of Wilson's Lattice
  Gauge Theories},''
\href{http://dx.doi.org/10.1103/PhysRevD.11.395}{{\em Phys. Rev.} {\bfseries
  D11} (1975) 395--408}.
%%CITATION = PHRVA,D11,395;%%.

\bibitem{Jordan1928}
P.~Jordan and E.~Wigner, ``{\"U}ber das paulische {\"a}quivalenzverbot,''
  \href{http://dx.doi.org/10.1007/BF01331938}{{\em Zeitschrift f{\"u}r Physik}
  {\bfseries 47} no.~9, (Sep, 1928) 631--651}.
  \url{https://doi.org/10.1007/BF01331938}.

\bibitem{farhi2000quantum}
E.~Farhi, J.~Goldstone, S.~Gutmann, and M.~Sipser, ``Quantum computation by
  adiabatic evolution,''
  \href{http://arxiv.org/abs/quant-ph/0001106}{{\ttfamily
  arXiv:quant-ph/0001106}}.

\bibitem{Jansen2007}
S.~Jansen, M.-B. Ruskai, and R.~Seiler, ``Bounds for the adiabatic
  approximation with applications to quantum computation,''
  \href{http://dx.doi.org/10.1063/1.2798382}{{\em Journal of Mathematical
  Physics} {\bfseries 48} no.~10, (2007) 102111}.

\bibitem{Lloyd1073}
S.~Lloyd, ``Universal quantum simulators,''
  \href{http://dx.doi.org/10.1126/science.273.5278.1073}{{\em Science}
  {\bfseries 273} no.~5278, (1996) 1073--1078}.

\bibitem{Suzuki1991}
M.~Suzuki, ``General theory of fractal path integrals with applications to
  many‐body theories and statistical physics,''
  \href{http://dx.doi.org/10.1063/1.529425}{{\em Journal of Mathematical
  Physics} {\bfseries 32} no.~2, (1991) 400--407},
  \href{http://arxiv.org/abs/https://doi.org/10.1063/1.529425}{{\ttfamily
  https://doi.org/10.1063/1.529425}}.

\bibitem{Coleman:1976uz}
S.~R. Coleman, ``{More About the Massive Schwinger Model},''
\href{http://dx.doi.org/10.1016/0003-4916(76)90280-3}{{\em Annals Phys.}
  {\bfseries 101} (1976) 239}.
%%CITATION = APNYA,101,239;%%.

\bibitem{Byrnes:2002nv}
T.~Byrnes, P.~Sriganesh, R.~Bursill, and C.~Hamer, ``{Density matrix
  renormalization group approach to the massive Schwinger model},''
  \href{http://dx.doi.org/10.1103/PhysRevD.66.013002}{{\em Phys. Rev. D}
  {\bfseries 66} (2002) 013002},
  \href{http://arxiv.org/abs/hep-lat/0202014}{{\ttfamily
  arXiv:hep-lat/0202014}}.

\bibitem{Iso:1988zi}
S.~Iso and H.~Murayama, ``{Hamiltonian Formulation of the Schwinger Model:
  Nonconfinement and Screening of the Charge},''
  \href{http://dx.doi.org/10.1143/PTP.84.142}{{\em Prog. Theor. Phys.}
  {\bfseries 84} (1990) 142--163}.

\bibitem{Nielsen-Chuang}
M.~A. Nielsen and I.~L. Chuang, {\em Quantum Computation and Quantum
  Information: 10th Anniversary Edition}.
\newblock Cambridge University Press, USA, 10th~ed., 2011.

\bibitem{Buyens:2015tea}
B.~Buyens, J.~Haegeman, H.~Verschelde, F.~Verstraete, and K.~Van~Acoleyen,
  ``{Confinement and string breaking for QED$_2$ in the Hamiltonian picture},''
  \href{http://dx.doi.org/10.1103/PhysRevX.6.041040}{{\em Phys. Rev. X}
  {\bfseries 6} no.~4, (2016) 041040},
  \href{http://arxiv.org/abs/1509.00246}{{\ttfamily arXiv:1509.00246
  [hep-lat]}}.

\bibitem{Kogut:1981ny}
J.~B. Kogut, D.~K. Sinclair, R.~B. Pearson, J.~L. Richardson, and
  J.~Shigemitsu, ``{The Fluctuating String of Lattice Gauge Theory: The Heavy
  Quark Potential, the Restoration of Rotational Symmetry and Roughening},''
  \href{http://dx.doi.org/10.1103/PhysRevD.23.2945}{{\em Phys. Rev. D}
  {\bfseries 23} (1981) 2945}.

\bibitem{Pearson:1981nz}
R.~B. Pearson and J.~L. Richardson, ``{THE QUARK - ANTI-QUARK STATE IN THE
  STRONG COUPLING LIMIT OF LATTICE QCD},''
  \href{http://dx.doi.org/10.1103/PhysRevD.25.2658}{{\em Phys. Rev. D}
  {\bfseries 25} (1982) 2658}.

\bibitem{Wiebe_2012}
N.~Wiebe and N.~S. Babcock, ``Improved error-scaling for adiabatic quantum
  evolutions,'' \href{http://dx.doi.org/10.1088/1367-2630/14/1/013024}{{\em New
  Journal of Physics} {\bfseries 14} no.~1, (Jan, 2012) 013024}.
  \url{http://dx.doi.org/10.1088/1367-2630/14/1/013024}.

\bibitem{Shaw2020quantumalgorithms}
A.~F. Shaw, P.~Lougovski, J.~R. Stryker, and N.~Wiebe, ``Quantum {A}lgorithms
  for {S}imulating the {L}attice {S}chwinger {M}odel,''
  \href{http://dx.doi.org/10.22331/q-2020-08-10-306}{{\em {Quantum}} {\bfseries
  4} (Aug., 2020) 306}. \url{https://doi.org/10.22331/q-2020-08-10-306}.

\bibitem{Eastin2009}
B.~Eastin and E.~Knill, ``Restrictions on transversal encoded quantum gate
  sets,'' \href{http://dx.doi.org/10.1103/PhysRevLett.102.110502}{{\em Phys.
  Rev. Lett.} {\bfseries 102} (Mar, 2009) 110502}.
  \url{https://link.aps.org/doi/10.1103/PhysRevLett.102.110502}.

\bibitem{Bravyi2012}
S.~Bravyi and J.~Haah, ``Magic-state distillation with low overhead,''
  \href{http://dx.doi.org/10.1103/PhysRevA.86.052329}{{\em Phys. Rev. A}
  {\bfseries 86} (Nov, 2012) 052329}.
  \url{https://link.aps.org/doi/10.1103/PhysRevA.86.052329}.

\bibitem{Paetznick2014}
A.~Paetznick and K.~M. Svore, ``Repeat-until-success: Non-deterministic
  decomposition of single-qubit unitaries,'' {\em Quantum Info. Comput.}
  {\bfseries 14} no.~15–16, (Nov., 2014) 1277–1301.

\bibitem{Bocharov_2015}
A.~Bocharov, M.~Roetteler, and K.~M. Svore, ``Efficient synthesis of universal
  repeat-until-success quantum circuits,''
  \href{http://dx.doi.org/10.1103/physrevlett.114.080502}{{\em Physical Review
  Letters} {\bfseries 114} no.~8, (Feb, 2015) 080502}.

\bibitem{Tran_2021}
M.~C. Tran, Y.~Su, D.~Carney, and J.~M. Taylor, ``Faster digital quantum
  simulation by symmetry protection,''
  \href{http://dx.doi.org/10.1103/prxquantum.2.010323}{{\em PRX Quantum}
  {\bfseries 2} no.~1, (Feb, 2021) 010323}.

\bibitem{Adam:1997wt}
C.~Adam, ``{Massive Schwinger model within mass perturbation theory},''
  \href{http://dx.doi.org/10.1006/aphy.1997.5697}{{\em Annals Phys.} {\bfseries
  259} (1997) 1--63},
\href{http://arxiv.org/abs/hep-th/9704064}{{\ttfamily arXiv:hep-th/9704064
  [hep-th]}}.
%%CITATION = HEP-TH/9704064;%%.

\bibitem{Interval}
T.~Okuda.
\newblock In preparation.

\bibitem{Funcke:2019zna}
L.~Funcke, K.~Jansen, and S.~K\"uhn, ``{Topological vacuum structure of the
  Schwinger model with matrix product states},''
  \href{http://dx.doi.org/10.1103/PhysRevD.101.054507}{{\em Phys. Rev. D}
  {\bfseries 101} no.~5, (2020) 054507},
  \href{http://arxiv.org/abs/1908.00551}{{\ttfamily arXiv:1908.00551
  [hep-lat]}}.

\end{thebibliography}\endgroup
%%%%%%%%%%%%%%%%%%%%%%%
%%%%%%%%%%%%%%%%%%%%%%%
%%%%%%%%%%%%%%%%%%%%%%%

\end{document}